\documentclass[a4paper,11pt,headings=big,DIV=12]{scrartcl}
\usepackage{amsmath,amssymb}
\usepackage[colorlinks,linkcolor=mygray,urlcolor=blue,citecolor=blue]{hyperref}
\usepackage{tcolorbox} 
\definecolor{mygray}{gray}{0.2}
\definecolor{mypink1}{rgb}{0.9, 0.2, 0.6}
\usepackage{graphicx}
\usepackage{enumerate} 
\usepackage{slashed}
\usepackage{cite} 
\usepackage{comment} 
\usepackage{multirow}
 \usepackage{bbold}
\allowdisplaybreaks
\newcommand{\vev}[1]{\langle #1 \rangle} 
\newcommand{\state}[1]{|#1\rangle}

\newcommand{\matel}[3]{\langle #1|#2|#3\rangle}
\newcommand{\al}{\alpha}
\newcommand{\be}{\beta}
\newcommand{\ga}{\gamma}
\newcommand{\de}{\delta}
\newcommand{\la}{\lambda}
\newcommand{\eps}{\epsilon}

\newcommand{\MeV}{\,\mbox{MeV}}

\newcommand{\LaQCD}{\Lambda_{\textrm{QCD}}}

\newcommand{\epsIR}{\eps_{\text{IR}}}

\newcommand{\VCKM}[1]{V_{\textrm{#1}}}

\newcommand{\claa}{c_{\AAA}}

\newcommand{\AAA}{a } 
\newcommand{\qSaa}{q^2_{\AAA}}
\newcommand{\cl}{c_{\ell}}

\newcommand{\clz}{c_{0}}

\newcommand{\Cdot}{\!\cdot \!}
\newcommand{\mi}{\!-\!}
\newcommand{\pl}{\!+\!}

\newcommand{\TAB}{Tab.~}
\newcommand{\FIG}{Fig.~}
\newcommand{\SEC}{Sec.~}
\newcommand{\SECs}{Secs.~}
\newcommand{\APP}{App.~}
\newcommand{\APPs}{Apps.~}

\newcommand{\Rea}{\textrm{Re}}
\newcommand{\Ima}{\textrm{Im}}

\newcommand{\mga}{m_\gamma}

\newcommand{\ORD}{{\cal O}}

\newcommand{\DEL}{\delta}

\newcommand{\Amp}{{\cal A}}
\newcommand{\Ampzero}{{\cal A}^{(0)}}
\newcommand{\Ampone}{{\cal A}^{(1)}}
\newcommand{\Amptwo}{{\cal A}^{(2)}}

\newcommand{\lone}{\ell_1}
\newcommand{\ltwo}{\ell_2}

\newcommand{\pK}{p_\pi}
\newcommand{\pB}{p_B}

\newcommand{\inclusive}{e^+ e^- \to hadrons}
\newcommand{\leptonic}{\pi^+ \to \ell^+ \bar \nu}
\newcommand{\semileptonic}{B \to \pi \ell^+  \bar \nu}

\newcommand{\emphB}[1]{\textbf{ #1}}

\newcommand{\ppi}{p}

\newcommand{\ONE}{i) }
\newcommand{\TWO}{ii) }

\newcommand{\CVA}{C_{V\mi A}}
\newcommand{\CSP}{C_{S\mi P}}

\newcommand*{\mathcolor}{}
\def\mathcolor#1#{\mathcoloraux{#1}}
\newcommand*{\mathcoloraux}[3]{%
  \protect\leavevmode
  \begingroup
    \color#1{#2}#3%
  \endgroup
}

\begin{document}

%\begin{flushright}
%\begin{tabular}{l}
%%CP3-Origins-2016-042 DNRF90 \\
% DIAS-2016-4
%CP3-Origins-2016-042 DNRF90
%\end{tabular}
%\end{flushright}
%\vskip1.5cm

\begin{center}
{\Large\bfseries \boldmath Notes on QED Corrections in Weak Decays}
\\[0.8 cm]
{\Large%
 Roman Zwicky
\\[0.5 cm]
\small
 Higgs Centre for Theoretical Physics, School of Physics and Astronomy,\\
University of Edinburgh, Edinburgh EH9 3JZ, Scotland 
} \\[0.5 cm]
\small
E-Mail:
\texttt{\href{mailto:roman.zwicky@ed.ac.uk}{roman.zwicky@ed.ac.uk}}.
\end{center}

\bigskip
\pagestyle{empty}

\begin{abstract}\noindent
In these lecture notes the basics of QED corrections to hadronic decays are reviewed 
with special emphasis on conceptual (e.g. counting and tracking of infrared sensitive logs) 
rather than numerical aspects. 
General matters are illustrated for the cases of increased  complexity and decreased 
inclusiveness:  $e^+ e^- \to hadrons$,  the leptonic decay  $\pi^+ \to \ell^+  \bar \nu$ and 
the semileptonic decay  $B \to \pi \ell^+  \bar \nu$. The  non-trivial and ongoing efforts of 
including structure dependence are very briefly outlined.
 \end{abstract}

\setcounter{tocdepth}{3}
\setcounter{page}{1}
\tableofcontents
\pagestyle{plain}

\section{Introduction}

Quantum electrodynamics (QED) can be regarded as the oldest and possibly most accurate and successful quantum field theory (QFT) there is.  The renormalisation of QED, 
by the pioneers Dyson, Feynman, Schwinger, Tomonaga and others \cite{Schweber:1994qa}, gave birth 
to the successful application of quantum field theory to all of particle physics culminating
in the Standard Model (SM) in the sixties \cite{Glashow:1961tr,Weinberg:1967tq,Salam:1968rm} and  finally the Higgs-boson discovery    in 2012 
\cite{ATLAS:2012yve,CMS:2012qbp}.  Since the QED coupling constant is small $\alpha  \equiv \frac{e^2}{4 \pi} \approx \frac{1}{137}$ perturbation theory is a reliable  tool for  many cases. A topical example  is 
the  anomalous magnetic moment of the muon $a_\mu = (g_\mu-2)/2$
with the theory average  $a_\mu =116591810(43) 10^{-11}$ \cite{Aoyama:2020ynm}
very close to the  experimental average 
$a_\mu =116592061(41) 10^{-11}$ 
\cite{Muong-2:2021ojo}, currently with some tension.

The application of QED to particle decays comes with additional subtleties 
which can be traced back to two idealisations, infinite space and infinitely precise measurement apparatuses, which do not hold in practice leading to infrared- (IR) divergences.   In well-defined observables IR-divergences 
cancel and  the understanding thereof is based on cancellation-theorems 
\cite{Bloch:1937pw,Kinoshita:1962ur,Lee:1964is}  relying on first principles such as  unitarity.
IR-sensitivity, leading to large logs,   can invalidate the naive counting in perturbation theory.  
In  $d \Gamma( B \to \pi e^+ \bar \nu)/d E_\pi$ for example, one will find   
$\alpha \to \alpha  \ln m_b/m_e \approx 0.05$ to all orders in perturbation theory. 
The focus on these notes is on conceptual matters of QED in weak decays illustrated on examples.
In the remaining two paragraphs we briefly comment on important topics 
not covered in this text.

In reporting experimental results in flavour physics the QED-radiation is regarded 
as a background and is effectively removed 
by using Monte-Carlo programs such as PHOTOS \cite{PHOTOS} or PHOTONS++ in SHERPA \cite{Schonherr:2008av}. 
These tools are based on versions of scalar QED (point-like approximations). 
The cross-validation of these programs seems essential in assuring  precision 
extraction of CKM matrix elements (e.g. $|\VCKM{u(c)b}|$) or the 
testing of  lepton flavour universality \cite{Bifani:2018zmi} 
(e.g. $R_K =  \Gamma[B  \to K \mu^+ \mu^-]/\Gamma[B \to K  e^+ e^-]$ with tensions since 2014 up to 
its latest measurement \cite{LHCb:2021lvy} ).
This topic certainly deserves further commenting and study.\footnote{Let us add that one needs to distinguish kaon physics from $D$- and $B$-physics in this respect. In the former case the situation is better as the logs are not that large, structure-dependent analyses in chiral perturbation theory exists 
and experiment is more inclusive  in the photon such that Monte-Carlo tools are 
not indispensable in principle.}  Somewhat related QED is also important in the context of initial state radiation in $e^+ e^-$ colliders \cite{Frixione:2022ofv} and the main proponent in QED in strong 
backgrounds \cite{Fedotov:2022ely}.

We will not review the infrared problems of quantum chromodynamics (QCD) but refer the reader to 
an excellent list of text books \cite{Sterman:1993hfp,Weinberg:1995mt,Muta:1998vi,Smilga:2001ck,Collins:2011zzd}
and review articles \cite{Sterman:1995fz,Sterman:2004pd,Agarwal:2021ais}. We content ourselves emphasising
that QCD is conceptually very different from QED in that there is a mass gap for the observable hadronic spectrum. 
All particle masses are proportional to 
a non-perturbative scale $\LaQCD = \ORD( 200\MeV)$ with the exception 
of the pseudo-Goldstone, due to chiral symmetry breaking,  for which $m_\pi^2 = m_q \ORD(\LaQCD)$.
 The challenge in QCD is to
establish factorisation theorems whereby collinear divergences arising from a hard kernel, computed with quarks and gluons,  
are absorbed in a meaningful way into hadronic objects such as the parton distribution 
functions or  jets.  

%This resembles some problems of structure dependence and QED correction discussed at the end of this small review.

 These short notes  are organised as follows. 
 In \SEC\ref{sec:IR} we describe the origin of infrared divergences and the cancellation thereof in observables. Three examples,  $\inclusive$, $\leptonic$ and $\semileptonic$  in increasing complexity are reviewed in \SEC\ref{sec:examples} at the level of the point-like approximation.  Aspects of going beyond this approximation are discussed in \SEC\ref{sec:beyond} and we end with conclusions in \SEC\ref{sec:conclusions}. 
 Formal matters such as the Low-theorem, the KLN-theorem and coherent states 
 are summarised or extended in \APPs \ref{app:Low}, \ref{app:KLN} and  \ref{app:coherent}.
 Some more practical aspects related to QED, such as 
  infrared singularities at one-loop, numerical handling of singularities   and terminology  
  can be found in \APPs  \ref{app:regions}, \ref{app:pragmatic} 
 and \ref{app:terminology}  respectively.

%The order of the section titles is: Introduction, Materials and Methods, Results, Discussion, Conclusions for these journals: aerospace,algorithms,antibodies,antioxidants,atmosphere,axioms,biomedicines,carbon,crystals,designs,diagnostics,environments,fermentation,fluids,forests,fractalfract,informatics,information,inventions,jfmk,jrfm,lubricants,neonatalscreening,neuroglia,particles,pharmaceutics,polymers,processes,technologies,viruses,vision

\section{Infrared Divergences and Infrared-sensitivity}
\label{sec:IR}

IR-divergences are associated with massless particles and there 
are two known mechanisms for enforcing massless particles, 
Goldstone bosons and gauge bosons 
(without confinement and unbroken gauge symmetry).\footnote{The fermion mass in QCD can be put to zero and remains zero in perturbation theory
due to chiral symmetry  but the zero value in itself does not stand out by any mechanism.}$^,$\footnote{ 
Not so long ago it has been understood that the photon can be viewed as a Goldstone boson of a higher form symmetry \cite{Gaiotto:2014kfa}. 
This would bring down the number of mechanisms  to one and further unify the picture. }
The Goldstone effective theory, chiral perturbation theory in QCD, is largely free 
from IR-divergences as   the shift symmetry enforces derivative interactions which 
tame the IR-behaviour. 
Now, the only gauge boson of the type 
mentioned is our well-known photon and this places  QED as a unique laboratory for 
IR-problems.\footnote{To some extent this also applies to the graviton and gravity as 
already studied by Weinberg  \cite{Weinberg:1965nx} and \cite{Cachazo:2014fwa} for renewed interest.}
Before venturing any  further it is advisable   to review  the basics of
 IR-divergences.  Since real and virtual photon radiation 
are connected by cancellation theorems  it is  sufficient, at first, to consider 
real radiation only.  
\begin{figure}[h]
\begin{centering}
\includegraphics[width=7.0 cm]{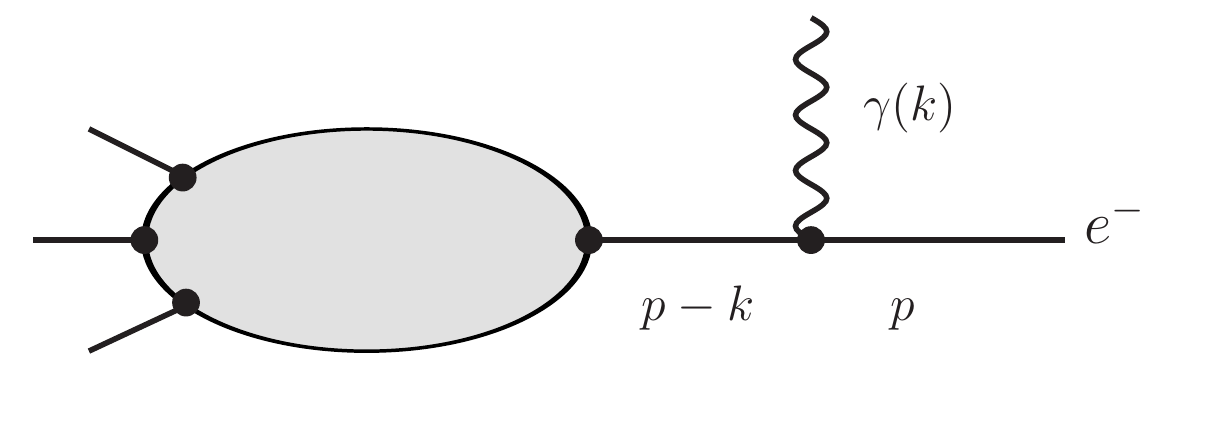}
\caption{\small Photon-emission from an external electron in a generic process. 
\label{fig:real-emission}}
\end{centering}
\end{figure} 
Disregarding ultraviolet (UV) divergences the only type of divergences that can arise  
in Feynman diagrams are due to  propagators going on-shell which are of the IR-type.  At LO this is particularly 
simple as we may just consider 
real emission of a photon from a charged particle, e.g. a lepton such as the electron $e^-$, as depicted in \FIG\ref{fig:real-emission}.   
The propagator $\frac{1}{(p+k)^2 - m_e^2}$ denominator, for on-shell $p$, behaves like
\begin{equation}
\label{eq:basic}
(p+k)^2 - m_e^2  = 2 p \cdot k  = 2 E_\ga E_e( 1 - \be \cos \theta) \;,
\end{equation}
where $k = E_\ga(1,0,0,1)$, $p = (E_e,  \kappa \hat{n} )$, 
 $E_e = \sqrt{m_e^2 + \kappa^2}$,  $\be = \kappa/E_e$ and $\theta$ the angle between 
 the unit vector $\hat{n}$ and the $z$-axis. The propagator is singular if either the photon energy $E_\ga $ or the angle $\theta$ approach zero 
(and $m_e \to 0$). These divergences are known as  
soft and collinear  respectively. In $d=4$ they lead to logarithmic singularities $\ln m_\ga$ and $\ln m_e$.\footnote{A photon mass $m_\ga$  is introduced to regulate the soft divergence, in addition to \eqref{eq:basic},
which in dimensional regularisation would map
into $\frac{1}{\epsIR}$. Note that the photon mass also regularises the collinear divergences.} In certain regions of phase space these divergences combine 
and lead to soft-collinear divergences $\ln m_\ga \, \ln m_e$.  
Generally, at $n$-loops there are terms of the order $\ln^k m_\ga \, \ln^l m_e$ with 
$l \leq n$ and $k+l \leq 2n$.

 It seems worthwhile to
briefly digress on the collinear term $\ln m_e$. For finite lepton mass this is a physical 
effect, see  for example the previously mentioned  sizeable
$\al \ln m_e/m_b$-terms  in $B \to \pi e^+  \bar \nu$.\footnote{
In QCD $\ln m_q$ terms are either absorbed into hadronic quantities such as distribution amplitudes, parton distribution functions or jets in the context of what is known as \emph{factorisation theorems} 
or if this  cannot be done then the variable is not IR-safe. 
This might indicate a problem of applying perturbation in a non-perturbative regime.}
The question for what observables QED is well-defined for zero lepton masses, gave rise 
to the KLN-theorem (cf. \APP\ref{app:KLN} for further comments). 
We shall assume leptons masses to be non-zero and 
special emphasis will be given to $\ln m_\ell$-terms to which 
we refer to as \textbf{ hard-collinear logs} (cf. also \APP\ref{app:terminology}) and are a physical effect.  
This contrasts the terms caused by zero energy photons to which refer to as 
 IR-divergences (and interchangeably as soft-divergences) following the main literature.

\subsection{Observables are infrared finite}

Of course physical observables have to be free of divergences and this is where 
one expects deep physical principles to dictate cancellations. 
Cancellations segregate  observable from 
non-observable quantities. First,  IR-divergences are interlinked with 
the very definition of what a particle is and the measurement process itself. 
How can one distinguish a single electron from an electron with an ultrasoft photon
(or a highly relativistic electron with a photon emitted at an infinitesimally small angle)? That is also indeed 
where the resolution lies, what is measurable needs to be assessed carefully. 
One needs to come back to the idealisation mentioned in the introduction: infinite space and infinite detector resolution. 
The  true IR-divergences (i.e. excluding collinear ones)  are 
effectively regulated by the introduction of an energy scale, say $\DEL$ (which has to be larger 
than the actual detector resolution scale). There are two main approaches to it:
\begin{enumerate}
\item 
The fixed particle Fock-space is abandoned in favour of so-called \textbf{ coherent states}  which 
take into account that charged particles  are surrounded by a  soft photon-cloud 
\cite{Chung:1965zza,Kibble:1968oug,Kibble:1968npb,Kibble:1968lka,Kulish:1970ut}.
In 1970 Kulish  and Faddeev \cite{Kulish:1970ut}  showed that the coherent state approach leads  to a finite $S$-matrix in QED, which is gauge invariant with a separable Hilbert space.\footnote{There is no successful version of this approach for perturbative QCD, 
for early attempts see  \cite{Giavarini:1987ts} 
and for recent improved  understanding of the underlying issues thereof cf. \cite{Gonzo:2019fai}. 
From a purely conceptual viewpoint this is not crucial as the $S$-matrix of QCD is defined with respect to its
physical states, the  hadronic states, and it comes with all its good properties.  The $S$-matrix elements
can be extracted from (non-perturbative) correlation functions via the LSZ-formalism (as shown to be valid 
by the Haad-Ruelle scattering theory  \cite{Duncan}).
On a pragmatic level, in collider physics, quarks and gluons hadronise into jets.} 
\item Second,   one defines  \textbf{ observables} which are \textbf{ inclusive enough}  such that these divergences cancel.  This approach was pioneered by Bloch and Nordsieck 1937 \cite{Bloch:1937pw}, extended in the sixties by the KLN-theorem \cite{Kinoshita:1962ur,Lee:1964is} (cf. \APP\ref{app:KLN})
to additionally include collinear singularities  and applied to correlation functions in form of the Kinoshita-Poggio-Quinn-theorem  \cite{Kinoshita:1962ur,Poggio:1976qr,Sterman:1976jh} (cf. \SEC\ref{sec:inclusive}).
As a rule of thumb, the more inclusive a quantity is, the fewer divergences or IR-sensitive terms there are.
 \end{enumerate}
The second approach can, reassuringly,  be seen as a limit of the  latter.
In view of it being more general we consider it worthwhile to first discuss the coherent 
state approach.  Our brief summary  is largely   based 
on the excellent presentation in Duncan's book \cite{Duncan} and some more context can be found in \APP\ref{app:coherent}.
Let us concretely assume that the detector can only capture 
photons with an energy above $ \DEL$ and reject photons with energies above that threshold.
Thus it is advisable to replace 
the electron state, to which we adhere for illustration,  by a state with any number of photons with energies smaller than the detector cut-off
\begin{equation}
\label{eq:coherent}
\state{e^-(\vec{q})} \to  \state{e^-(\vec{q})}_n \equiv \state{e^-(q),\ga(k_1) \dots \ga(k_n) }_{(E_\ga)_i < \DEL} \;,
\end{equation}
and (formally)
\begin{equation}
\state{e^-(\vec{q})}  = \sum_{n \geq 0,\vec{q}} c_n(\vec{q}) \state{e^-(\vec{q})}_n \;,
\end{equation}
is the coherent state, with appropriate $c_n(\vec{q})$, which can be written as an exponential of an  integral over the creation operators
cf. \APP\ref{app:coherent}. 
Denoting by $P_n$ the probability of $n$-soft photon  emission, the total 
probability is a sum of all possibilities 
$ P_{\textrm{tot}} = \sum_{n \geq 0} P_n$.   
When the total transition probability $P_{\textrm{tot}}$
of all  $n$-states \eqref{eq:coherent} is considered, 
the momentum space integrals are cut-off below at $\DEL$ and are thus manifestly IR-finite (no soft-divergences).  The $S$-matrix is well-defined, as mentioned above, and  the IR-divergences 
are absorbed into the definition of the states. 
It seems worthwhile to point out that this bears some resemblance with the absorption of the UV divergences into the parameters of the theory which in turn 
also originates from an idealisation, namely that space-time is a continuum.  

How does this connect to the Bloch-Nordsieck mechanism?
Reassuringly,  upon expanding to finite order in $\al$ one recovers 
the Bloch-Nordsieck solution. 
More concretely, in order to compute the $\ORD(\al)$ corrections to  
a decay  process $i \to f$ one has to consider its radiative counterpart $i \to f (\ga_{E_\ga < \DEL})$. In the total transition probability one can show that the IR-divergences cancel diagram by diagram; as beautifully illustrated in many textbooks e.g. 
 $e^+ e^- \to \bar q q $ in \cite{Muta:1998vi}. These cancellations have been shown to hold 
to all orders in QED by exponentiation \cite{Yennie:1961ad,Weinberg:1965nx}.\footnote{The case of QCD, which is beyond the scope of these notes,
 is complicated as the simple combinatorics in QED are spoiled by 
zero mass charged particles (the gluons) and 
the colour structure. 
The Bloch-Nordsieck mechanism is replaced in perturbation theory by the KLN-theorem, 
whose features are briefly  discussed in \APP\ref{app:KLN}, and for the more involved case of hadrons in final states we refer to the textbooks \cite{Sterman:1993hfp,Collins:2011zzd}.}

\paragraph{No  fixed particle-number $S$-matrix:}
Let us  briefly digress and motivate why 
the $S$-matrix of the fixed particle-number Fock space does not exist, as it turns out to be zero. 
We provide three different viewpoints:
\begin{enumerate}
\item 
The IR-divergences, caused by  the absence of a mass gap, can be seen as 
an indication  for the ill-defined fixed particle number Fock space $S$-matrix. 
It is instructive to give mass to the particle causing the trouble, the photon. 
The IR-divergences $\ln m_\ga$  exponentiate such that the $S$-matrix, 
$S \propto \exp( |a| \ln m_\ga + \dots) \to 0 $, assumes zero value 
 in the limit of zero photon mass. 
 Hence, the $S$-matrix is infinite at fixed order and zero at all orders!
 Thus the asymptotic completeness 
of the in and out Hilbert space ceases to make sense as there is no $S$-matrix connecting the two.
\item Another way to look at it is to realise that due to the massless photons the single particle pole, assumed by the LSZ-formalism, is softened by the presence of radiative corrections 
$(p^2-m^2)^{-1} \to (p^2-m^2)^{-1 + \al |A| }$ into a branch cut 
as first shown by Schroer in 2D model  \cite{Schroer:1963gw} (he came up with  the term ``infraparticle").
This makes the particle of mass  $m$ disappear from the $S$-matrix when multiplied by the LSZ-factor 
$p^2 - m^2$ upon taking the on-shell limit $p^2 \to m^2$. 
Moreover,  Buchholz  \cite{Buchholz:1986uj} has shown, using very general arguments,  that a charged particle obeying Gauss' law cannot be a discrete eigenstate of the momentum squared operator, which goes hand in hand with the branch cut. A notable aspect is that the coefficient $|A|$ is gauge dependent, 
e.g. \cite{Jackiw:1968jpv}, and another sign that there is a problem.  
\item  
Whereas the $S$-matrix is gauge invariant in perturbation theory this is not the entire story 
as has recently been shown using  asymptotic symmetries \cite{Kapec:2017tkm}. 
The common lore is that local gauge symmetries   give rise to global charge conservation only 
and that local gauge symmetries are not really symmetries in the observable sense. 
However, asymptotically (that is at spatial infinity) 
there are infinitely many symmetries, so-called asymptotic 
symmetries \cite{Strominger:2017zoo}, known as large (i.e. non-local)  gauge symmetries.  
In a very interesting paper \cite{Kapec:2017tkm} it has been shown that the vanishing of the fixed particle number $S$-matrix can be understood as due to  non-invariance under these asymptotic symmetries. Closing the circle, it is found that once gauge invariance is enforced, the coherent states emerge! 
 % It is tempting to speculate that the non-gauge invariance of the exponent $A$ in point 2 and the large gauge symmetries in point 3 are connected.  
\end{enumerate}
Now, is it considered a problem that the fixed  particle number $S$-matrix is not defined?  
For mathematical physics, yes.  The fact that electron does not correspond to an isolated particle in the spectrum 
is known as the IR-problem of QED  (e.g. 
  \cite{Mund:2021zhx} also for historic references and 
discussion of this notorious problem). 
The pragmatic particle physicist, or
advocate of the Bloch-Nordsieck- and KLN-approach, would simply point to 
the fact that  the fixed  particle number $S$-matrix is not  an observable but rather an intermediate auxiliary quantity.  
%However, as stated above, a scattering theory can be formulated between 
%the coherent states of the type \eqref{eq:coherent} with a well-defined $S$-matrix \cite{Kulish:1970ut}.
 
Concluding,  in practice the infrared problem of QED is bypassed in the pragmatic 
approach by IR-regularisation (e.g. $m_\ga \neq 0$) 
and removing the regulator ($m_\ga \to 0$)  in 
observables such as decay rates. 
Let us add that in practice, for a number of reasons (e.g. no additional scale), dimensional regularisation  is the choice by most practitioners.

%%%%%%%%%%%%%%%%%%%%%%%%%%%%%%%%%%%%%%%%%%
\section{Decay Rates and their Infrared-effects}
\label{sec:examples}

Following the discussion on the origin of IR-divergences and 
why they disappear from observables we discuss these mechanisms  in three practical examples
 with decreasing level of inclusiveness and increasing level of IR-effects. 
 Namely, 
the (inclusive) $e^+ e^- \to hadrons$ cross section, the leptonic decay $\leptonic$
and  the semileptonic case $\semileptonic$. 
In the latter two cases, the hadrons 
will be treated in the point-like approximation with comments beyond this treatment deferred to \SEC\ref{sec:beyond}.

For most practical applications first order $\ORD(\al)$ is sufficient.  
At the amplitude level we therefore need $\ORD(e^{0,1,2,})$, denoted by  
$\Amp^{(0,1,2)}$, corresponding to tree, real and virtual. 
We refer to  $\Ampzero$ and $\Amptwo$ as the non-radiative  and to $\Ampone$ as the 
radiative amplitude. 
The cancellation of IR-divergences is then a result of $\Rea[\Ampzero (\Amptwo)^*]$ versus $|\Ampone|^2$ when properly integrated over phase space. 
Let us rephrase this in terms of a generic decay $i \to f$ at the level of the rates
\begin{alignat}{3}
\label{eq:IRtype}
&d \Gamma(i \to f) &\;\propto\;&   1 +& &  \frac{\al}{\pi}(A_V \ln m_\ga +  B_V  \ln  m_{\ga,f} \ln m_f 
+ C_V \ln m_f  + \ORD(1)  ) d \Phi_f   \;, \nonumber \\[0.1cm] 
&d \Gamma(i \to f \ga) &\;\propto\;&  & &  \frac{\al}{\pi}(A_R \ln m_\ga +  B_R  \ln m_{\ga,f} \ln m_f 
+ C_R \ln m_f  + \ORD(1)  ) d \Phi_{f} d\Phi_\ga  \;,
\end{alignat}
where $d \Phi$ is the phase space measure,  
$m_f$ is a small mass of a final state particle  (e.g. an electron mass)
and $m_{\ga,f}$ stands for either  $m_f$ or $m_\ga$. The subscripts $V$ and $R$ denote  virtual and real and $A$, $B$ and $C$ stand 
for soft, soft-collinear and hard-collinear divergences. 
Integrating over the entire photon phase space 
\begin{eqnarray}
\label{eq:diff-cancel}
d \Gamma(i \to f)  + \int d\Phi_\ga \,  d \Gamma(i \to f\ga )   \propto 
 (1 + \frac{\al}{\pi} C  \ln m_f + \ORD(1) ) \, d \Phi_f \;,
\end{eqnarray}
with all the soft-divergences canceling and the collinear logs cancel,
\begin{equation}
C = ( C_V + \int d\Phi_\ga C_R)  =  
\left\{\begin{array}{ll} 
\textrm{zero} & \textrm{collinear-safe differential variables} \\[0.1cm]
\textrm{non-zero} & \textrm{non collinear -safe  differential variables} 
\end{array} \right. \;,
\end{equation}
depending on the differential variables (cf. \SEC\ref{sec:semileptonic} for a concrete example).
The further statement of the  cancellation-theorems (Bloch-Nordsieck and KLN)  is that that if one integrates over the remaining  phase space $d \Phi_{f}$,  then (in the total rate) 
\begin{equation}
\label{eq:total-cancel}
\Gamma(i \to f) + \Gamma(i \to f\ga)  \propto  1 + \frac{\al}{\pi}\ORD(1) \;,
\end{equation}
all IR-divergences are absent, schematically: $([A,B,C]_V+ [A,B,C]_R)^{\textrm{inc}} = 0$. 
This picture is  broken in practice by the following two sources:
\begin{itemize}
\item [\ONE] The experiment is not fully photon-inclusive 
and rejects hard photons with $E_\ga > \DEL$ where $\DEL$ is 
the previously discussed threshold which is (slightly) larger than the actual detector resolution.\footnote{If one of the final state particles is very light then one might think to apply cuts on the angle because of the angular resolution as well.  As long as the mass of the charged particle is finite one can separate 
it from the collinear photon(s)  by a magnetic field.} 
 This leads to the replacements
\begin{alignat}{2}
\label{eq:IRtypeII}
& (A_V+ A_R) \ln m_\ga &\;\to\;&  (A_V(\DEL)+ A_R(\DEL)) \ln \DEL  \;,\nonumber \\[0.1cm] 
& (B_V+ B_R) \ln m_\ga \ln m_f &\;\to\;&  (B_V(\DEL)+ B_R(\DEL)) \ln \DEL \ln m_f \;,  \nonumber \\[0.1cm] 
& C \ln m_f &\;\to\;& C(\DEL) \ln m_f \;,
\end{alignat}
where  $C(\DEL) \neq 0$  irrespective of whether the differential  variables are collinear-safe 
or not. The functions $A,B,C(\DEL)$ are polynomial in $\DEL$.
\item [\TWO] The rate can be  differential in some final state kinematics and therefore not a total rate as in 
\eqref{eq:diff-cancel}. 
In this case the unitarity argument, on which the cancellation is based, does not 
necessarily hold since the kinematics make the sum too restrictive.
The (non)-cancellation needs to be reassessed, depending on the kinematic variables 
hard-collinear effects  $\ln m_f$ do or do not cancel. 
\end{itemize}

\begin{table}[h] 
%%% \tablesize{} %% You can specify the fontsize here, e.g., \tablesize{\footnotesize}. If commented out \small will be used.
\begin{center}
\begin{tabular}{l |  l | l | l  | r}
\textbf{type}	& \textbf{\ONE diff. in $\ga$}	& \textbf{\TWO diff. in $f$} &  \textbf{IR-terms} &  \textbf{\SEC}\\
\hline
$\inclusive	$	& no			& no & none & \ref{sec:inclusive}  \\[0.1cm]
$\leptonic	$	& yes			& no & $A,B, (C)$ Eqs.\mbox{(\ref{eq:IRtype},\ref{eq:IRtypeII})}  & \ref{sec:leptonic}   \\[0.1cm]
$\semileptonic	$	& yes			& yes & $A,B,C$ Eqs.\mbox{(\ref{eq:IRtype},\ref{eq:IRtypeII})} & \ref{sec:semileptonic}   \\[0.1cm]
\end{tabular}
\end{center}
\caption{Types of observables considered where diff. is short for differential in $\ga$ or 
$f$ (final states) and \ONE and \TWO refer to the itemised conditions above. The bracket around (C) 
in row 2 will, hopefully, become clear upon reading \SEC\ref{sec:leptonic}.
 \label{tab:cases}}
\end{table}

\subsection{A classic example of infrared finiteness: $\inclusive$}
\label{sec:inclusive}

Here we briefly deviate from the QED-course as we consider finiteness 
under correction in the strong coupling constant to $\inclusive$. 
An analogue in QED 
would be the somewhat exotic $ \nu \bar \nu\to Z \to \ell^+ \ell^-$.  
Now, by the optical theorem the total cross-section 
\begin{equation}
\label{eq:e+e-}
\sigma_{\textrm{tot}} (\inclusive)(q^2) \propto \Ima [  \Pi(q^2) ]  \;,
\end{equation}
is related to the imaginary part of 
 the vacuum polarisation $\Pi(q^2)$
\begin{equation}
\label{eq:vacpol}
\left(q_\mu q_\nu - q^2  g_{\mu\nu}   \right){\Pi}(q^2)   =  
 i \int d^4 x  e^{i x \cdot q}  \matel{0}{T j_\mu (x) j_\nu(0)  }{0}   \;,
\end{equation}
where $j_\mu = \sum_f e Q_f \bar f \ga_\mu f$  is the electromagnetic current 
and $Q_f$ the electromagnetic charge.
On the non-perturabative level  there is no question as to whether this quantity is well-defined because of the mass gap. In particular, in the  large-$N_c$ limit
\begin{equation}
\Ima [  \Pi(s) ] =  \pi \sum_{V = \rho^0,\omega ..}
\delta(s-m_V^2)  f_V^2 \;,
\end{equation}
 with $f_V$ the vector meson decay constants and most importantly $m_{\rho^0} \approx 770\MeV$ is the lowest mass exhibiting the mass gap.  
The question we would like to address is whether it is finite to all orders in perturbation theory using  quarks and gluons as degrees of freedom.  

According to the cancellation-theorems and the discussion outlined in the beginning 
of this section this must be the case since this is a  fully inclusive observable 
(and conditions \ONE \& \TWO are not met).
Alternatively, this can be established on grounds of the Kinoshita-Poggio-Quinn-theorem  
\cite{Kinoshita:1962ur,Poggio:1976qr,Sterman:1976jh} which states: 
\emph{In massless renormalisable theories the one-particle irreducible correlation functions are IR-finite for non-exceptional 
(external) Euclidean momenta.}\footnote{Non-exceptional momenta configurations are such that no subset of momenta adds to zero.}  
Renormalisability is important as it settles power counting for the proof and the Euclidean momenta condition avoids  particles going on-shell. This applies to the case at hand since 
$\Ima [  \Pi(q^2) ] = \frac{1}{2i}( \Pi(q^2 + i0) -  \Pi(q^2 - i0))$ with $q^2 \pm i0$ effectively counts as off-shell (or Euclidean in practice). Hence $\sigma_{\textrm{tot}}(q^2)$ must be IR-finite (in perturbation theory)  as found in many explicit computations for any $q^2 > 0$ in particular. 
\begin{figure}[h]
\begin{centering}
\includegraphics[width=10.0 cm]{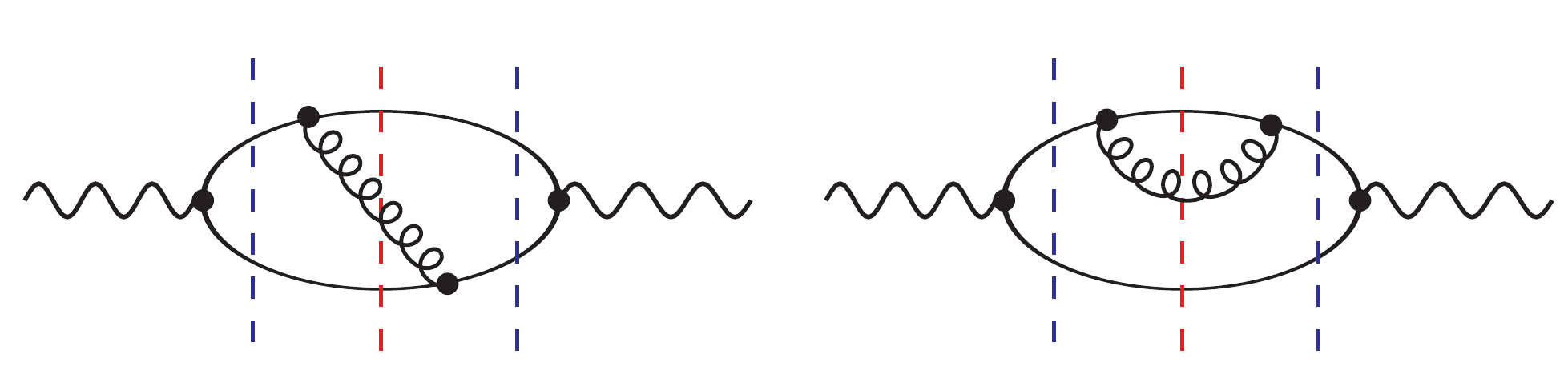}
\caption{\small Strong coupling corrections  to the vacuum polarisation $\Pi(q^2)$ \eqref{eq:vacpol}, at $O(\al_s)$, 
which necessarily involves quarks and gluons (partons).  As its imaginary part corresponds to the total cross section \eqref{eq:e+e-} the cuts give rise to various subprocesses which include the virtual and real parts. The dashed or blue cuts  correspond to the virtual and the real parts respectively. 
\label{fig:real-emission}}
\end{centering}
\end{figure}  
One can learn a fair amount by  considering  
the one-loop corrections (depicted in  \FIG\ref{fig:real-emission}) 
since the imaginary part is proportional to the discontinuity
and the latter is proportional to the sum of all cuts by the Cutkosky rules (e.g. \cite{Sterman:1993hfp}) 
The different types of cuts include the radiative and non-radiative parts cf. figure caption. Each one 
of these cuts is IR-divergent but they cancel in the sum as dictated by the arguments given above. That individual contributions behave very different from the total contribution is not restricted to IR-effects but can also appear in the power-behaviour of a heavy quark mass or an external momentum in case they are assumed to be large.

\subsection{Leptonic decay of the type $\leptonic$}
\label{sec:leptonic}

We now turn to the  simple example of an exclusive decay,   
the pion  decay $\leptonic$.   The photon energy cut-off $E_\ga < \DEL$ (in say the  pion restframe)  will introduce the $\ln \DEL$-terms 
as in \eqref{eq:IRtypeII}. 
This will lead to soft- and soft-collinear terms 
as indicated in \TAB\ref{tab:cases}. 
The hard-collinear logs ($C$-type in \eqref{eq:IRtype}) are a bit peculiar 
in this decay in the SM  since the amplitude is
 $\ORD(m_\ell )$
(and therefore automatically finite in the limit $m_\ell \to 0$).  
This helicity suppression is relieved for $S\mi P$ interactions and we thus include them along  the $V\mi A$ structure in order to illustrate the straightforward nature of the hard-collinear logs in this example. 
In turn these logs have  to disappear in the photon-inclusive limit $2 m_\pi  \DEL \to m_\pi^2 - m_\ell^2$.
All of which will be made explicit.

The four-Fermi  effective Lagrangian,  including $S\mi P$- and $V\mi A$-interactions,
reads 
\begin{equation}
\label{eq:Leff}
{\cal L}^{\textrm{eff}} =  4 \sqrt{2} G_F\left(   \CVA \bar u \ga_\mu d_L \bar{\ell} \ga^\mu \nu_L  +  \CSP  \bar u  d_L \bar{\ell}  \nu_L \right)\;,
\end{equation}
where $2 f_L \equiv (1-\ga_5) f$ and in the  SM $(\CVA,\CSP)  =( \VCKM{ud},0)$.
The LO amplitude  is given by
\begin{alignat}{2}
\label{eq:AmpLepLO}
& \Ampzero(\leptonic)  &\;\propto\; &   \CVA (L_0)_\mu  H^\mu _0   +  \CSP L_0 H_0    \nonumber \\[0.1cm]
& &\;=\; &  i (  \CVA  m_\ell   F_\pi  - \CSP G_\pi    )  L_0   \;,
\end{alignat}
where  the leptonic matrix element reads 
\begin{equation}
L_0^{(\mu)} \equiv \matel{\bar{ \nu} \ell^+ }{\bar \ell \Gamma^{(\mu)} \nu } {0}  = 
   \bar u(p_\nu) \Gamma^{(\mu)}  \nu(p_\ell) \;,
\end{equation}
with  $\Gamma = (1-\ga_5)$,  $ \Gamma_\mu  =  \ga_\mu \Gamma$
and the
hadronic matrix elements are 
\begin{alignat}{3}
\label{eq:Fpi}
&  \matel{0} {A^a_{5 \, \mu } }{ \pi^b (p)}  &\;=\;& \de^{ab} (H_0)_{\mu} &\;=\;& 
\phantom{-}i  \de^{ab} F_{\pi} \ppi_{\mu}  \;, \nonumber \\[0.1cm]  
&   \matel{0} {P^a }{ \pi^b (p)} &\;=\;& 
 \de^{ab} H_0 &\;=\;&  - i  \de^{ab} G_{\pi}  \;, \quad G_{\pi}  = \frac{F_{\pi} m_\pi^2}{2 m_q} = \frac{- \vev{\bar qq}}{2 F_\pi} \;,
\end{alignat}
with $A^a_{5 \, \mu}   =  \bar q T^a \gamma_{\mu} \gamma_{5} \, q$, 
$P^a   =  \bar q T^a  \gamma_{5} \, q$  and $T^a$ the 
adjoint $SU(2)$-representation matrix corresponding to  $q = (u,d)$ (with (u)p and (d)own quarks).  Note that use of the equation of motion was made for the $V\mi A$-part 
in \eqref{eq:AmpLepLO} which makes the  $m_\ell$-suppression factor 
 explicit.
The LO decay rate  is given by 
\begin{equation}
 \Gamma(\leptonic)^{(0)} = \frac{G_F^2 }{ \pi m_\pi^3}   | \CVA  m_\ell F_\pi   - \CSP G_\pi  |^2    |\vec{p}_\ell|^2  \;,
\end{equation}
where the lepton velocity, in the pion's restframe, is
\begin{equation}
\quad |\vec{p}_\ell|  = 
\frac{\la^{1/2}(m_\pi^2,m_\ell^2,0)}{2 m_\pi} = 
\frac{m_\pi}{2}  \left( 1- \frac{m^2_\ell}{m_\pi^2} \right) \;,
\end{equation}
and  $\la(x,y,z) \equiv x^2+y^2+z^2 - 2x y -2 x z -2 y z$ denotes  the K\"all\`en function.
Notably $F_\pi \approx 92 \MeV$, a non-perturbative parameter of QCD known 
as the pion decay constant,  is the order parameter of  
the spontaneous 
breaking of chiral symmetry $SU(2)_L \times SU(2)_R \to SU(2)_V$ 
(in the $m_q  \to 0$ limit).
When QED corrections are considered it ceases to be an observable and 
it is essentially degraded to the status of a wave function renormalisation constant. 
This can be seen from the explicit results in the nice review \cite{Gasser:2010wz}
where $F_\pi$ is found to be gauge dependent and divergent  in the $m_\pi \to 0$  limit.
Unlike in QCD, in QED  the chiral logs $\ln m_\pi $  are not protected 
by powers in the pion mass  since $F_\pi$ is not an observable. This is a point we will come back to at the end of the section. 
\begin{figure}[h]
\begin{centering}
\includegraphics[width=10.0 cm]{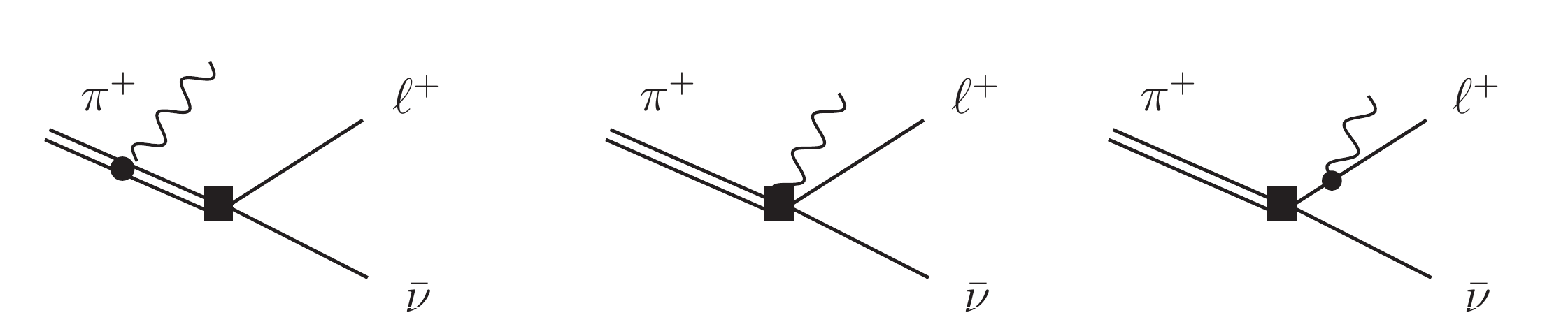}
\caption{\small Real emission diagram of the pion decay. The  diagram in the centre is the so-called contact term and does appear for the $V\mi A$- but not the $S\mi P$-interaction. 
The real amplitude is given in \eqref{eq:leptonic-real}. 
\label{fig:leptonic}}
\end{centering}
\end{figure} 
Next we discuss how to incorporate radiative corrections in the point-like approximation. 
%This has first been undertaken by Kinoshita and improved by Marciano and Sirlin in computing the matching to the weak scale.  
This is a straightforward exercise in effective field theory. 
The hadronic operator are matched to pions ($\matel{0}{\pi^a}{\pi^b} = \de^{ab}$)
\begin{equation}
A^a_{5 \, \mu}  \to - F_\pi D_\mu  \pi^a \;, \quad P^a  \to - i G_\pi  \pi^a \;,
\end{equation}
such that  the LO matrix 
element \eqref{eq:Fpi} is reproduced. 
The momentum dependence  in the axial current \eqref{eq:Fpi} enforces
a covariant derivate, $D_\mu \pi^a = (\partial_\mu + i e Q_{\pi^a} A_\mu) \pi^a$, which gives raise to a so-called contact term. 
The leading radiative amplitude is given by 
\begin{alignat}{2}
\label{eq:leptonic-real}
& \Ampone(\leptonic \ga) \propto
\sum_{i } C_i  & & \left( \hat{Q}_{\ell_1}  \bar u \frac{2 \eps^* \cdot 
\hat{\ell}_1+ \slashed{\eps}^* \slashed{k} }{ 2 k \cdot 
\hat{\ell}_1  }  (\Gamma \cdot H_0)_i  v + 
\hat{Q}_{\bar{\ell}_2}  \bar u   (\Gamma \cdot H_0)_i  \frac{2 \eps^* \cdot 
\hat{\ell}_2+ \slashed{k} \slashed{\eps}^*  }{2  k \cdot 
\hat{\ell}_2  } v  \; + \right.  \nonumber \\[0.1cm]
& & &  \left.   \hat{Q}_{\pi} (L_0 \cdot H_0)_i|_{\ppi \to \bar{\ppi}} \frac{\eps \cdot 
(\hat{p} + \hat{\bar{p}} )  }{ 2 k \cdot 
\hat{p}  } 
 +   \hat{Q}_{\pi} (L_0 \cdot H_0)_i|_{\ppi \to \eps^*}  \right) \;,
\end{alignat}
where $\bar{p} = p - k$,  
$ \ell,\nu \to \ell_1 ,\ell_2$  in order to be more general, 
$i = S \mi P, V \mi A$  and the conventions are the same as in \cite{Isidori:2020acz}: 
  $\hat{Q}_j  = \pm Q_j$  and $\hat{p}_j  = \pm p_j$ for out(in)-going states.  Some more detail on the notation. 
  The last term in \eqref{eq:leptonic-real}, and centre of \FIG\ref{fig:leptonic}, 
 is the so-called contact term,  only present for  $V\mi A$  as mentioned above.
 In addition, the following compact notation has been  introduced 
\begin{equation}
 (L_0 \cdot H_0)_i =   \left\{\begin{array}{ll} 
(L_0)_\mu  H^\mu _0   & i = V \mi A
    \\[0.1cm]
L_0  H _0   & i = S \mi P    \end{array} \right. \;,
\end{equation}
 likewise for $L_0 \to \Gamma$.  The terms of the Low-theorem (cf. \APP\ref{app:Low}) 
 are explicit which include the $\ORD(E_\ga^{-1})$ eikonal terms 
 \begin{equation}
 \label{eq:eikonal}
  \Ampone = \Ampzero  \sum_i \hat{Q}_i \frac{\eps^* \cdot \hat{p}_i}{k \cdot   \hat{p}_i} + \ORD(E_\ga^0) \;,
 \end{equation}
  and the $\ORD(E_\ga^0)$-term related to the angular momentum can be seen in the leptonic parts.
Gauge invariance amounts to $\Ampone|_{\eps \to k} = 0$ and does hold provided
$\sum_i \hat{Q}_i =   \hat{Q}_{\ell_1}  + \hat{Q}_{\bar{\ell}_2} +  \hat{Q}_{\pi}  = 0$ (which is nothing but charge conservation).  The latter has to be imposed in gauge-fixed perturbation theory but would be automatic in a manifestly gauge invariant formalism such 
as the path-integral used in lattice simulations. 
Hence the radiative amplitude is  gauge invariant and thus the   virtual (or non-radiative) amplitude 
  must be as well.  In particular in the virtual amplitude the gauge dependence of the $\ORD(\al)$ pion decay constant cancels against the lepton-pion 
and lepton  radiative corrections.\footnote{In fact in the virtual case one finds 
that the covariant gauge-fixing parameter $\xi$ appears in the form
$\Amptwo \propto \xi (\sum_i \hat{Q}_i)^2 + \dots$ and is again effectively absent because of charge conservation  \cite{Isidori:2020acz}. This time the charge condition is quadratic of course.}
As previously said, we present  the $S\mi P$- and $V\mi A$-interaction separately as they both have different features.

\subsubsection{Leading logs with $S\mi P$-interaction}
\label{sec:SP}

For the  $S\mi P$-interaction ($\CSP \neq 0, \, \CVA=0$)  we may parameterise the $\ORD(\al)$ rate as follows\footnote{Soft logs are proportional to the LO rate but not hard-collinear which arise in differential distribution (cf. \eqref{eq:magic} in the next section).}
\begin{equation}
\label{eq:leptonic-rate}
 \Gamma(\leptonic) =    \Gamma(\leptonic)^{(0)}(1 + \frac{\al}{4 \pi}\left(F_{\textrm{soft}}(\hat{m}^2_\ell , 2 \hat{\DEL})   + F_{\textrm{coll}}(\hat{\DEL}) \ln \hat{m}_\ell  + \textrm{non-log}   \right)) \;,
 \end{equation}
 where ``non-log" stands for anything that is neither a soft, soft-collinear or hard-collinear log.
 Hatted quantities, except charges,  are understood to be divided by the pion mass in this section. 
 The quantity  $\DEL  $  is the previously introduced photon energy cut-off and its photon-inclusive limit is $2 \hat{\DEL}  \to 1-\hat{m}_\ell^2$.
 Below we discuss both $F_{\textrm{soft}}$ and $F_{\textrm{coll}}$ without resorting to the full computation.
 \begin{itemize}
 \item \emph{The soft and soft-collinear terms} are universal  and given by 
 \begin{equation}
 \label{eq:soft}
F_{\textrm{soft}}(x,y) =  - (4 \frac{1+x^2}{1-x^2}  \ln x^2 + 8)  \ln y \;,
 \end{equation}
 and  its exponentiation is a well established \cite{Yennie:1961ad,Weinberg:1965nx} 
 \begin{equation}
 \label{eq:resum}
 \Gamma( \al \to \be) = \Gamma( \al \to \be)^{\textrm{LO}} \exp (  - A \ln  \frac{\la}{\Lambda}) \;, \quad 
 \end{equation}
 where $\la$ and $\Lambda$ are IR and UV cut-offs.  
 These are to be replaced in practice with $ \DEL$ and 
 the largest scale in the problem; beyond that they are equivalent 
 to so-called finite terms and  undetermined in the leading log 
 approximation.\footnote{We will have more to say on how this happens in computation  in \SEC\ref{sec:semileptonic}. 
 The breaking of Lorentz-invariance by introducing a photon energy cut-off in a specific frame introduces 
 a practical challenge.}
Now,  the factor $A$ has a pleasing form
 \begin{equation}
 \label{eq:A}
 A =  \frac{e^2}{8 \pi^2} \sum_{i,j} \hat{Q}_i\hat{Q}_j  \frac{1}{2 \be_{ij}} \ln \frac{1+\be_{ij}}{1-\be_{ij}} \;,
 \end{equation}
 where the sum is over the charged particles in the decay and 
\begin{equation}
\be_{ij} = \frac{ \be_i + \be_j}{1+\be_i \be_j} \;,
\end{equation}
 is the relativistic addition of the velocities of the $i,j$-particles in the $ij$-restframe.
 With  $\be_{ii}= 1$   for $i = \pi^+,\ell^+$ (since  the relative velocities are zero 
) and with $\be_{\ell \pi} =  (1 - \hat{m}_\ell^2)/(1+\hat{m}_\ell^2)$ 
 one recovers $\eqref{eq:soft}$.
 
 It is instructive to reproduce the leading term  from  the eikonal part \eqref{eq:eikonal} which is of course
 what the original papers did. 
Following \cite{Isidori:2020acz}  we denote the decay rate as 
\begin{alignat}{2}
\label{eq:schema}
& d \Gamma &\;=\;& d \Gamma^{\textrm{LO}} + \frac{\al}{\pi} \sum_{i,j} \hat{Q}_i \hat{Q}_j (
{\cal H}_{ij} + {\cal  F}_{ij}(\DEL) ) d\phi_f =  d \Gamma^{\textrm{LO}}(1 + \Delta_{\text{rel}}  \, d \phi_f) \;,
\end{alignat}
where ${\cal H}$ and ${\cal F}$ stand for the non-radiative and the radiative part respectively 
and $\Delta_{\text{rel}} $ is the relative correction, not to be confused with the photon energy cut-off, which is a function of the non-trivial differential variables  $d \phi_f = \prod_{i=1}^{n_f} d\vartheta_i$  (with $n_f = 0$ and $n_f=2$  in the leptonic and semileptonic case respectively).
After making use of gauge invariance, by choosing the Feynman gauge $\xi =1$, performing the 
polarisation sum  $\sum_{\la} \eps^*_\mu(\la) \eps_\nu(\la)  =  - g_{\mu \nu} 
+ (1-\xi) k_\mu k_\nu/k^2 \to -g_{\mu\nu}$  over the eikonal part one gets
\begin{alignat}{1}
& {\cal F}_{ij}(\DEL) =  (2 \pi)^2 \int_{\DEL}  \frac{-  {p}_i \cdot {p}_j }{( k \cdot{p}_i )( k \cdot {p}_j )  }  d\Phi_\ga = - K_R(\DEL) I_{ij}^{(0)} + \textrm{non-soft}   \;,
 \label{eq:F02}
 \end{alignat}
where  ``non-soft" stands for finite non-logarithmic regularisation dependent terms.  
The $K_R(\DEL)$-term is the regularisation dependent energy integral and $ I_{ij}^{(0)}$ an angular integral.
In the leading log approximation $K_R(\DEL)$ and $ I_{ij}^{(0)}$ are separately Lorentz invariant \cite{Isidori:2020acz}.
 This is non-trivial  since the introduction of the 
photon energy cut-off introduces a preferred frame and complicates the analytic evaluation of the non-approximated integrals. 
More concretely, 
\begin{equation}
K_R(\DEL)   =\int _{0}^{\DEL}\frac{dE_{\gamma }}{E_{\gamma }} 
=  \left\{\begin{array}{ll} 
  -\frac{1}{2} \ln \frac{\mga}{\mu} +\ln \left ( \frac{ \DEL}{\mu } \right ) +\ORD(\mga  )  & m_\ga \textrm{-reg} \\  
 - \frac{1}{2\eps} +\ln \left ( \frac{2\DEL }{\mu } \right ) + \ORD(\eps ) & 
\text{dim-reg}   \end{array} \right.  \;,
\end{equation}
given in dimensional regularisation $d = 4 - 2 \eps$ and photon mass regularisation 
(cf. \APP D  \cite{Isidori:2020acz} for some more detail). 
The angular integral produces a term
\begin{equation}
\label{eq:soft-integral}
I_{ij}  =  \int d \Omega  \frac{E_\ga^2  p_i \cdot p_j}{ (k \cdot p_i ) (k \cdot p_j)} =  \frac{1}{2 \be_{ij}} \ln \frac{1+\be_{ij}}{1-\be_{ij}} = 1 + \ORD(\be_{ij}) \;,
\end{equation} 
which matches the expression in \eqref{eq:A} and thus reproduces \eqref{eq:soft} as outlined earlier.
 \item \emph{The hard-collinear logs} can be obtained from the splitting function which has been verified 
 in \cite{Isidori:2020acz}  for the more advanced semileptonic case.
 The formula for the collinear logs reads
 \begin{alignat}{2}
 \label{eq:collin}
& \Delta_{\text{rel}}|_{\ln m_\ell} &\;=\;&  -  \frac{\al}{\pi} \hat{Q}^2_{\ell^+}    \ln \hat{m}_\ell 
\left( \frac{d\Gamma^{\textrm{LO}}} {d \phi_f} \right)^{-1}  
\int^1_{z(\hat{\DEL})} d z  P_{f \to f \ga}(z)   \frac{d \Gamma^{\textrm{LO}} }{d \phi_f } ( f_i (z)\vartheta_i )
\nonumber \\[0.1cm]
 & &\;\to\;&   - \frac{\al}{\pi} \hat{Q}^2_{\ell^+}    \ln \hat{m}_\ell 
 \int^1_{1- 2 \hat{\DEL}} {d z}  P_{f \to f \ga}(z) \nonumber \\[0.1cm]
  &  &\;=\;&   - \frac{\al}{\pi} \hat{Q}^2_{\ell^+}    \ln \hat{m}_\ell 
 \left( \frac{3}{2} - 2 \hat{\DEL}(2 - \hat{\DEL} ) \right)\;,
 \end{alignat}
(and thus $F_{\textrm{coll}} (\hat{\DEL}) = -4 \hat{Q}^2_{\ell^+} ( \frac{3}{2} - 2 \hat{\DEL}(2 - \hat{\DEL} )$)   with fermion splitting function
 \begin{equation}
 \label{eq:P}
 P_{f \to f \ga}(z)  =   \frac{1 + z^2}{(1- z)_+}  + \frac{3}{2} \de(1-z) \;,
 \end{equation}
 where $\de(1-z)$ is a Dirac delta function and 
  $\frac{1}{(1-z)_+}$  is the plus distribution 
$\int_0^1 dz  \frac{f(z) }{(1-z)_+}= \int_0^1 dz \frac{ f(z)- f(1))}{1-z}$.\footnote{This is just one specific way to regularise. Alternatively one may use for instance ${P}_{f \to f \ga}(z)  = \lim_{z^* \to 0} \left[  \frac{1 + z^2}{(1- z)}\theta((1-z^*)-z)  +( \frac{3}{2} + 2 \ln z^*)\de(1-z)   \right] $.} 
For the leptonic case the formula is trivial since there are no phase space variables. 
 Crucially, in the photon-inclusive limit $2 \hat{\DEL} \to 1$ the hard-collinear logs cancel 
 $ F_{\textrm{coll}} (\frac{1}{2})  =0$
    in accordance with the KLN-theorem.  This has to hold since 
    $\int_0^1 dz P_{f \to f \ga}(z) = 0$ which in turn follows from the conservation of the electromagnetic current
    (as it is related to the current's anomalous dimension which vanishes). \end{itemize}

\subsubsection{Leading order result with $V\mi A$-interaction as in the Standard Model}
\label{sec:lepVA}

The Standard Model computation  ($\CSP =  0,\, \CVA \neq 0$)  has of course been obtained a long time ago  \cite{Kinoshita:1959ha,Marciano:1993sh}, we quote 
\begin{equation}
\label{eq:leptonic-rate}
 \Gamma(\leptonic) =    \Gamma(\leptonic)^{(0)}(1 + \frac{\al}{4 \pi}\left(- 3 \ln
 \hat{m}_W^2   + F(\hat{m}^2_\ell , 2 \hat{\DEL})   \right)) \;,
 \end{equation}
 and comment on the various terms further below.
 In \eqref{eq:leptonic-rate}  $- 3 \ln {\hat{m}_W^2}$ incorporates  the matching to the 
$M_W$-scale \cite{Marciano:1993sh}.
 The explicit radiative function $ F(x,y)$  is given by \cite{Carrasco:2015xwa}
 \begin{alignat}{2}
&  F(x,y) &\;=\;&   4 \frac{1+x^2}{1-x^2} Li_2(y)   +  \ln x^2  + \frac{2-10 x^2}{1-x^2} \ln x^2 - 4 \frac{1+x^2}{1-x^2} Li_2(1-x^2) -3 \nonumber \\[0.1cm]
&  &\;+\;&  \frac{3+ y^2 + 4 y(x^2-1)}{2(1-x^2)^2} \ln (1\mi y) + \frac{y(4 \mi y \mi x^2)}{(1-x^2)^2} \ln x^2 + \frac{ y(22-3 y - 28 x^2)}{2(1-x^2)^2}  \nonumber \\[0.1cm]
&  &\;+\;& F_{\textrm{soft}}(x,y)  \;.
 \end{alignat}
  In the photon-inclusive case, $ F_{\textrm{inc}}(x) \equiv F(x,1-x^2)$, the radiative function assumes the form 
 \begin{alignat}{2}
 \label{eq:F2}
&  F_{\textrm{inc}}(x) &\;=\;&  - 8 \ln(1 - x^2) - \frac{3 x^2}{(1-x^2)^2} \ln x^2 - 8 \frac{1+x^2}{1-x^2}  Li_2(1-x^2)  \nonumber \\[0.1cm]
& &\;+\;&  \frac{13 - 19x^2}{2(1-x^2)} + \frac{6 - 14 x^2 - 4 (1+x^2) \ln (1 \mi x^2)}{ 1-x^2} \ln x^2) \;.
 \end{alignat}
 Let us now turn our focus to the logs as in the previous section:
\begin{itemize}
\item \emph{The soft and soft-collinear terms} are universal and $F_{\textrm{soft}}(x,y) $ is indeed the same 
function as in \eqref{eq:soft}.
\item \emph{Hard-collinear logs}, of the type $\ln m_\ell$,  are  not present. 
The LO $V\mi A$-amplitude is $\ORD(m_\ell)$-suppressed. 
and this is enough to guarantee the absence of the latter at $\ORD(\al)$ which 
 can be seen as follows. 
In the real radiation rate the $\ln m_\ell$-terms
arise from the eikonal part \eqref{eq:eikonal}  which are proportional to the 
LO amplitude which is $\ORD(m_\ell)$ and thus the logs can be at worst of the 
form  $m^2_\ell \ln m_\ell^2$ in the rate. Since the $\ln m_\ell$-terms in 
the virtual and the real part of the $\ORD(\al)$ rate have to cancel the virtual rate cannot 
contain them either. We are to conclude that $\ORD(\al) m_\ell^2 \ln m_\ell$ are the leading logs of this type. Since the limit $m_\ell \to 0$ is not divergent these logs  do
not have to cancel contrary to the $S\mi P$-case. Inspection of  \eqref{eq:F2} shows that they do indeed \emph{not cancel} since $F  = - 3\hat{m}_\ell^2 \ln  \hat{m}_\ell^2+ \dots$.
It seems worthwhile to briefly pause and reflect. 
 If the ``naive" equation of motion, linking $V \mi A$ to $S \mi P$, 
 where to apply it would be possible to reuse $S\mi P$-computation from the one of $V\mi A$.  This holds for the real part but not the virtual part as in this case the photon in the derivative interaction is \emph{not} an external on-shell particle.  

The moral of the story is that collinear logs only cancel if they have to due to the principle of unitarity which underlies the KLN-theorem.
\item \emph{A different type of collinear log:} 
We may however turn the tables and consider the decay $\tau^- \to \pi^- \bar \nu$
and regard $\ln m_\pi$ as the collinear log. The amplitude which is identical to the one 
for the leptonic decay is not $\ORD(m_\pi)$-suppressed,  thus there will be $\ln m_\pi$ terms 
in the real and the virtual part of the rate and they have to cancel in the total rate. There are some differences in the integration over phase space for the radiative part but not for the relevant eikonal terms.
Inspecting \eqref{eq:F2},  taking the $1/x \to 0$ limit and 
adding the log in \eqref{eq:leptonic-rate}, one collects $\frac{\al}{4 \pi}(6 + 16 + 6 + 0 + 0  -28  ) \ln m_\pi = 0$ and it is seen that the logs do cancel as they have to!
\end{itemize}

\subsection{Semileptonic decay of the type $\semileptonic$ }
\label{sec:semileptonic}

\begin{figure}[h]
\begin{centering}
\includegraphics[width=7.0 cm]{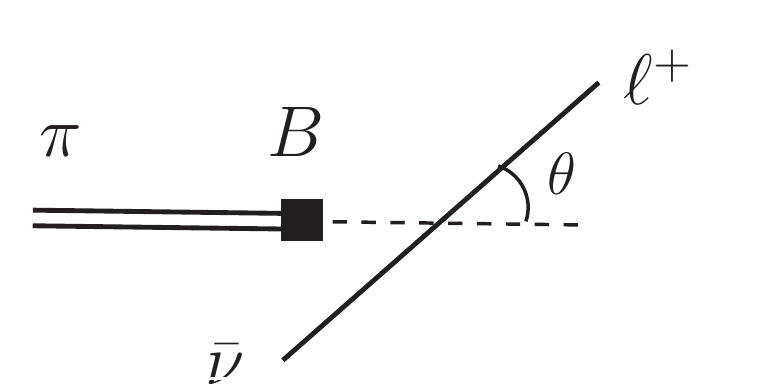}
\caption{\small Sketch of semileptonic (non-radiative) decay $\semileptonic$ with the definition of the 
lepton angle $\theta$ (and $q^2 = (p_{\ell^+} + p_\nu)^2$). The definition of these variables need to be revised when the photon emission is considered in addition \eqref{eq:tr}.}
 \label{fig:semileptonic}
\end{centering}
\end{figure}

The  new element in the semileptonic decay $\semileptonic$ is 
the extra meson in the final state  
leading to two non-trivial  kinematic variables. 
They  can be chosen to be the Dalitz-plot variables or the more commonly used lepton momentum squared 
$q^2 = (\ell_1+ \ell_2)^2$ and the angle $\theta$ of a lepton to the decay axis in the $q$-restframe 
(as depicted in \FIG\ref{fig:semileptonic}). Hence the LO decay  is differential unlike in the leptonic case 
(cf. for instance \APP B.1 in \cite{Isidori:2020acz} for the explicit result). 
A noticeable aspect is that QED, unlike the weak interaction term \eqref{eq:Leff}, give rise to higher 
moments in the lepton angle \cite{Gratrex:2015hna}. 

In many ways the QED-treatment of the  semileptonic decay $\semileptonic$ in the point-like approximation is similar to the leptonic decay and we shall be brief on those matters. 
There are also new aspects which bring in a certain amount of complication 
which we identify and examine more closely:
\begin{enumerate}
\item The role of the pion decay constant $F_\pi$ is taken by two form factors $f^{B \to \pi}_\pm(q^2)$,
\begin{alignat}{2}
&   \matel{0} {A^a_{5 \, \mu } }{ \pi^b (p)}  &\;=\;&  
-i  \de^{ab} F_{\pi} \ppi_{\mu}  \to  \nonumber \\[0.1cm] 
 & \matel{\pi }{ V_\mu }{ B }  &\;=\;&  f^{B \to \pi}_+(q^2) (p_B\pl p_\pi)_\mu +
  f^{B \to \pi}_-(q^2) (p_B\mi p_\pi)_\mu \;.
\end{alignat}
 Often in the literature 
the form factor is taken to be a constant, which is  a good approximation 
in $K \to \pi \ell^+ \bar \nu$ but less so for $\semileptonic$. 
Expanding the form factor in $q^2$, as in   \cite{Isidori:2020acz}, 
leads to a more involved  effective theory which goes beyond the point-like approximation. 
The  effect  of the expansion  is  most prominent when the photon energy cut is large 
due to migration of radiation 
(for which  we refer the reader to  the plots in \APP A  
in\cite{Isidori:2020acz}).\footnote{The flavour changing neutral current (FCNC) case is peculiar in that
for $B^0 \to \pi^0 \ell^+ \ell^-$  the form factor expansion amounts to the replacement of the constant form factor by $f_\pm \to f_\pm(q^2)$, whereas in the charged case $B^+ \to \pi^+ \ell^+ \ell^-$ the expansion is necessary and could be relevant because of the migration of radiation in conjunction with 
resonance-contributions entering non-resonant bins.} 
\item For the radiative matrix element  
the $(q^2 ,\theta)$-variables  have to be adapted because of the additional  photon. 
We follow the discussion in \cite{Isidori:2020acz} (replacing the kaon by the pion)  where 
the following kinematic variables
\begin{equation} 
\label{eq:tr}
\{ \qSaa,  \claa \}   = \left\{  \begin{array}{ll}
 q_\ell^2 = (\lone+\ltwo)^2\;, ~
 	&  \cl =  - \left(\frac{\vec{\lone}\cdot \vec{p}_{\pi}}{ |\vec{\lone}| | \vec{p}_{\pi}| } \right)_{q-\textrm{RF}}  \\
q^2_0 = (\pB-\pK)^2~, 
	&  \clz=  - \left(\frac{\vec{\lone}\cdot \vec{p}_{\pi}}{ |\vec{\lone}| | \vec{p}_{\pi}| } \right)_{q_0 -\textrm{RF}} \;,    \end{array} \right. 
\end{equation}
are defined with  $q-\textrm{RF}$ and $q_0-\textrm{RF}$ denoting  the  $q$ and 
$q_0 \equiv q+ k$ restframes respectively.  
Note that, the $(q_0^2,  \clz)$-variables, unlike at an $e^+ e^-$-collider, are difficult to measure at a hadron 
collider where the components of the $B$-momentum are unknown.
\item The LO amplitude is not  $\ORD(m_\ell)$-suppressed 
and a priori it is only the total (photon-inclusive) rate which is well-defined in the $m_\ell \to 0$  limit. 
As previously state, for finite $m_\ell$, as in the real world, this leads to a sizeable and measurable effect.
 This raises the interesting question as to whether  any of the differential variables 
 in \eqref{eq:tr} are collinear-safe (i.e. $m_\ell \to 0$ can be taken). 
\item The photon interacts with many particle-pairs and this complicates the analytic
evaluation of the phase-space integrals as one can choose the restframe only once. 
As previously discussed,  the energy- and  soft-integrals  \eqref{eq:soft-integral} are separately 
Lorentz-invariant in the soft-limit  and can therefore each  be evaluated using a separate preferred frame 
 \cite{Isidori:2020acz}.
\end{enumerate}
Now, point 4 is partly covered in \APP\ref{app:pragmatic} and all aspects of point 1 are covered in 
  \cite{Isidori:2020acz}.  
Let us just briefly mention that as long as a constant form factor is assumed 
or the mesons are neutral,  the computation of the real and virtual amplitude 
is very similar to the leptonic case albeit technically more involved.  
Points 2 and 3 deserve a closer look and are the topic of the next section.  

\subsubsection{(Non)-collinear safe differential variables}

 The soft-divergences which have to cancel at the differential level, can of course be derived using the same techniques as for the lepton case \eqref{eq:resum} (with relevant practical remarks deferred to \APP\ref{app:pragmatic}).
The hard-collinear divergences have been isolated using the phase space slicing technique.  They cancel charge by charge in the photon-inclusive total rate in accordance 
with \eqref{eq:total-cancel}. 
Let us now turn to the question, phrased in point 3, whether or not these logs cancel 
in one of the differential variables defined in \eqref{eq:tr}. It is found by explicit computation that the $\ln m_\ell$-terms cancel in the $(q_0^2,  \clz)$- but not the $(q^2,  \cl)$-variables \cite{Isidori:2020acz}.
 
 We wish to discuss this result from  a physics viewpoint. The 
 cancellation of soft-divergences  at the differential level   is quite plausible since 
 the soft photon does not make a difference to the radiative versus the non-radiative decay 
 topology.
 For the (energetic)  collinear photon  this is not the case.
 The topologies of the radiative and non-radiative amplitude are rather different 
 and a priori one would not expect cancellations. 
 In the total rate these cancellations are non-trivial and based on unitarity as emphasised earlier.  
  Thus it is natural to ask whether it can be understood from this viewpoint. 
 The answer is affirmative. 
 \begin{itemize}
\item  The $q_0^2$-variable is the four momentum of the total 
 lepton-photon system and for  fixed  $q_0^2$ one may interpret it as a decay 
 of a boson of mass $q_0^2$ into the two leptons and the photon, e.g. $W^+(q_0) \to \ell^+ \bar \nu (\ga)$.
And decay is not differential (its non-radiative part), just as the leptonic case, 
 and thus the $\ln m_\ell$ terms have to cancel by virtue of the KLN-theorem. 
 \item Alternatively one may regard $q_0^2$ as the analogue of a jet where 
 radiative and non-radiative parts are not distinguished and the problem of discerning 
 the lepton from the  lepton with a photon emitted at an infinitesimally small angle  does not pose itself.
 This is the pedestrian version of the IR-safety criteria which states that an observable $\Phi_n$ of 
 $n$-particles  is collinear-safe if (e.g. \cite{Contopanagos:1996nh})
 \begin{alignat}{1}
&   \Phi_n(p_1, \dots p_i,p_j, \dots p_n) \to  \Phi_{n-1}(p_1, \dots p_{ij}, \dots p_n)  \;,  \quad p_i \parallel p_j \;, \quad p_{ij} = p_i+p_j \;,
 \end{alignat}
 is smooth. Clearly this is the case in the $q_0^2$-variable but not the $q^2$-variable when differential.  
 \end{itemize}  
 \paragraph{Cancellation of hard-collinear logs in total rate:}
It is instructive to illustrate the cancellation of the hard-collinear terms in the total rate. 
Applying formula  \eqref{eq:collin} to the case where we keep the differential variable $q^2$ one gets
\begin{equation}
\frac{d\Gamma}{d q^2} |_{\ln \hat{m}_\ell}  = 
-  \frac{\al}{\pi} \hat{Q}^2_{\ell^+}  \ln \hat{m}_\ell 
 \int^1_{q^2/m_B^2} \frac{d z}{z}  P_{f \to f \ga}(z)   \frac{d \Gamma^{\textrm{LO}} }{dq^2 } ( q^2/z) +\ORD(1) \;,
\end{equation}
where the the factor $1/z$ is a Jacobian from the change of variable  $q_0^2 = q^2/z$ (with $z$ the energy fraction of the lepton after collinear splitting). 
The lower integration boundary of $z$ is the photon inclusive limit, neglecting $\ORD(m_{\pi,\ell})$-terms.
If we perform the integration over the $q^2$ phase space then the $\ln \hat{m}_\ell $-terms have to cancel 
according to the KLN-theorem.  This is indeed the case 
\begin{alignat}{2}
\label{eq:magic}
&\Gamma^{\textrm{tot}} |_{\ln \hat{m}_\ell} =  \int_{0}^{m_B^2} dq^2 \frac{d\Gamma}{d q^2} |_{\ln \hat{m}_\ell}  &\;\propto\;&  \int_0^1 \frac{dz}{z} 
\int_0^{z m_B^2} dq^2 P_{f \to f \ga}(z) \frac{d\Gamma^{\textrm{LO}}}{d q^2}(q^2/z)  \nonumber \\[0.1cm]
& &\;=\;&  \int_0^1 dz P_{f \to f \ga}(z) \int_0^{ m_B^2} d q_0^2   \frac{d\Gamma^{\textrm{LO}}}{d q^2}(q_0^2) =0  \;,
\end{alignat}
where in the first equation the order of integration has been  exchanged and in the second equation 
the chance of variable $q_0^2 = q^2/z$ was performed. This is of course the collinear-safe variable 
$q_0^2$ indeed. The vanishing of $ \int_0^1 dz P_{f \to f \ga}(z) = 0$ has been previously discussed in 
\eqref{eq:P}. In conclusion the hard-collinear logs vanish for the full rate independent of the specific decay rate. 
The assumption is of course that the splitting function reproduces all the logs. This fails if the $m_\ell \to 0$ limit can be taken such as in the leptonic decay of the SM (cf. \SEC\ref{sec:lepVA})
where the amplitude and the leading logs are
 $\ORD(m_\ell)$ suppressed.

%%%%%%%%%%%%%%%%%%%%%%%%%%%%%%%%%%%%%%%%%%
\section{Structure-dependent QED corrections - Resolving the Hadrons}
\label{sec:beyond}

\subsection{Summary on status of structure-dependent QED corrections} 
\label{sec:SDcorr}

The field of QED corrections to hadronic decays including structure-dependent corrections
(i.e. going beyond the point-like approximation) is not yet at a mature stage. 
The physical picture is well-motivated from the hydrogen atom where the proton and electron make up a charge neutral object but photonic interaction plays an important role. 
Thus it cannot be expected that a photon does not interact with a neutral $B$-meson 
composed of a $b$- and a $d$-valence quark. It is precisely for this meson that one can expect the largest effects  
as it is composed of a heavy and a light quark. 
There are various reasons why this is a difficult task. One of them is of course the cancellations of IR-divergences which enforces to consider real radiation.  
A task which goes beyond standard flavour physics and interferes with confinement 
at long distances. 

Amongst the continuum methods there  is chiral perturbation theory,  light-cone approaches such as soft-collinear effective theory (SCET) and heavy quark symmetry.  QED in chiral perturbation is well established  \cite{Cetal01,CGH08,Descotes-Genon:2005wrq},
and its  main challenge is the determination of the counterterms (which seem to follow the pattern of vector resonance saturation as in QCD). 
In SCET the leptonic FCNC decay $B_s \to \mu^+ \mu^-$ has been investigated in 
 \cite{BBS17,Beneke:2019slt} with the main parametric uncertainty coming from the 
 QCD $B$-meson distribution amplitude. Hadronic decays of the type 
 $B \to K \pi$ have been investigated in \cite{Beneke:2020vnb} and the definition of 
 the charged light-meson distribution amplitudes is non-trivial \cite{Beneke:2019slt}.  A remarkable aspect is that 
 so far in SCET only virtual contributions have been considered.  Real radiation
 is only  incorporated via the universal  soft-photon part \eqref{eq:resum}.  
 And heavy quark symmetry has been found to be constraining in $B \to D^{(*)} \ell \nu (\ga)$ 
 decays \cite{Papucci:2021ztr} (in the appropriate kinematic region).
 
 Lattice QCD + QED comes with its own challenges such as containing the massless photon in a finite box 
 (cf. \cite{Patella:2017fgk} for a review). There are, by now, four main programs. 
 QED$_L$ where the  spatial zero mode of the photon, which is in tension with  the finite volume,  
 is removed at the cost of locality 
 and non-gauge invariant interpolating operators are used for the charged mesons \cite{Carrasco:2015xwa}. 
 In this approach finite volume effects to hadronic observables (hadron masses and leptonic decay rates) are power-like rather than exponentially suppressed. In the context of leptonic decays, the leading universal finite volume effects have been determined up to $O(1/L)$ in \cite{Lubicz:2016xro} and up to $O(1/L^2)$ in Ref. \cite{DiCarlo:2021apt}, including structure-dependent contributions. Only virtual corrections are computed on the lattice and for the real correction the point-like approximation is proposed which is a good enough approach for $K^+,\pi^+ \to \ell^+ \nu$. First lattice results have been reported in \cite{Giusti:2017dwk,DiCarlo:2019thl} for these decays. A modification of this approach is QED$_\infty$ where finite volume effects are exponentially suppressed \cite{Feng:2018qpx}. This approach needs to be adopted case by case and has been applied 
 to the pion mass difference \cite{Feng:2021zek}.
Another approach is to work with a massive photon, emulating the continuum approach, which does not require to cut out the zero mode but introduces another scale into the problem \cite{Endres:2015gda}. First results
on hadron masses have been reported in \cite{Clark:2022wjy}.
A fully gauge invariant approach to lattice QCD, building upon ideas of Dirac and others, has been proposed 
\cite{Lucini:2015hfa}, known as $C^*$ boundary condition. Again results on  hadron mass determination have been reported \cite{Hansen:2018zre}.
%Decays have so far not been addressed in neither of the two last  this formalism.  

\subsection{Cancellation of hard-collinear logs for structure dependent contribution}
\label{sec:noSDlogs}

Technicalities aside, one may in particular be concerned that hard-collinear logs $\ORD(\al) \ln m_e/m_b$,
originating from structure-dependent corrections, do lead to large  
uncertainties as currently unknown.
Fortunately a rigorous result can be established forbidding those 
logs \cite{Isidori:2020acz},    based on gauge invariance. 
The basic idea of the proof is that when one considers a
light particle like the electron and  photon then $\ell_e = k + \ORD(m_e^2)$ in the collinear region which lends 
itself to the use of gauge invariance.  
We will sketch some more detail by decomposing the radiative amplitude ($\Ampone \to {\cal A}$ for brevity)  
\begin{equation}
{\cal A} = \eps^* \Cdot (  {\cal A}_e + ({\cal A} -  {\cal A}_e)) \;, \qquad  \eps^* \Cdot {\cal A}_e  \propto \hat{Q}_{e} \frac{\eps^* \cdot \lone}{k \cdot \ell_e} \;,
\end{equation}
such that the entire eikonal term of the electron is  in ${\cal A}_e$.  Squaring this matrix element, 
summing over polarisation in the Feynman gauge (cf. \SEC\ref{sec:SP}) 
and integrating over the photon phase space one gets three terms 
\begin{equation}
\int d \Phi_\ga \,  {\cal A}\Cdot {\cal A}^* = \int d \Phi_\ga ( ({\cal A} -  {\cal A}_e)\Cdot ({\cal A} -  {\cal A}_e)^* + 
2 \textrm{Re}[{\cal A}_e \Cdot   {\cal A}^*] -   {\cal A}_e \Cdot {\cal A}^*_e) \;.
\end{equation}
The first is by construction finite in the collinear region of the lepton $\lone$. 
The second has no hard collinear logs since it is proportional to 
\begin{equation}
\ell_e \cdot {\cal A}  = k \cdot {\cal A}   +  \ORD(m_{e}^2) = \ORD(m_{e}^2) \;, 
\end{equation}
 in the collinear region. 
The third one gives raise to the collinear logs. 
Firstly, we learn that the $\ln m_{e}$-terms  are necessarily proportional to $\hat{Q}_{e}^2$ 
(as manfiested in the splitting function approach).
Second, and more importantly  there cannot be any further hard collinear logs in the structure dependent part.  This is the case since the addition of structure dependent term will just change ${\cal A} \to {\cal A} + \de {\cal A}$ where $\de {\cal A}$ is itself gauge invariant 
and will be finite in the first term and not change the conclusions in the second either and not be part of 
the third one! The result is unchanged when the spin is considered, as explicitly shown for spin-$1/2$ and
argued for any spin in  \cite{Isidori:2020acz}.

Hence the result is: \emph{any gauge invariant addition (to the point-like approximation) can at most lead to logs of the form $\ORD(\al) m_e^2 \ln m_e$.} 
These terms are not sizeable and in particular  vanish in the chiral limit $m_e \to 0$.   
This result has been verified in  the derivative expansion of the form factor which 
is a particular approach that goes beyond the point-like approximation. 
This is fortunate as it puts $R_K$, or more generally  tests in the lepton universality, on much firmer grounds 
 since Monte Carlo tools such as PHOTOS do not (yet) incorporate structure-dependence.

% The MDPI table float is called specialtable

%%%%%%%%%%%%%%%%%%%%%%%%%%%%%%%%%%%%%%%%%%
\section{Discussions and Conclusions}
\label{sec:conclusions}

QED corrections have a long history. In particular electromagnetic corrections  
have been  the vehicle to the development of quantum mechanics and  QFT.  The massless photon leads to IR-effects which have a high degree of universality. 
The Bloch-Nordsieck cancellation mechanism from 1937, predates the solid development of QED in the 1940's, and is a strong indication of  universality in the IR-domain.  The IR-effects are interlinked with the measurement process and  gives rise to the largest QED corrections. 

We have reviewed the very basic of IR-divergences in \SEC\ref{sec:IR} along 
with the connection to the  elegant coherent states formalism.  
How IR-effects affect predictions was the topic of
\SEC\ref{sec:examples}, including three examples of increasing IR-sensitivity: 
the (inclusive) $e^+ e^- \to hadrons$ cross section, the leptonic decay $\leptonic$
and  the semileptonic case $\semileptonic$ in \SECs \ref{sec:inclusive}, \ref{sec:leptonic} and \ref{sec:semileptonic} respectively.   
 We have highlighted 
the peculiarity of the leading collinear logs in the leptonic decay in the Standard Model and clarified 
the importance of the choice of kinematic variables in the differential distribution of the 
semileptonic decay types.  Going beyond the point-like approximation, taking into account structure dependence, 
is the next step in the precision physics program of weak decays and a topic in \SEC\ref{sec:beyond}. 
Different methods and approaches have briefly been discussed in  \SEC\ref{sec:SDcorr}.
The text ends in \SEC\ref{sec:noSDlogs} with the model-independent demonstration, based on gauge invariance,  that the structure dependent part does not lead to new hard-collinear logs. 
This is fortunate as it will considerably reduce the uncertainty in many important observables such as 
the precision determination of heavy-light CKM-elements and tests of lepton flavour universality.
However, the implementation of these corrections in experiment 
will necessitate the development or extension of Monte Carlo tools. This demands a joint effort of theory
and experiment.

%%%%%%%%%%%%%%%%%%%%%%%%%%%%%%%%%%%%%%%%%%

%%%%%%%%%%%%%%%%%%%%%%%%%%%%%%%%%%%%%%%%%%
\vspace{6pt} 

%%%%%%%%%%%%%%%%%%%%%%%%%%%%%%%%%%%%%%%%%%
%% optional
%\supplementary{The following are available online at \linksupplementary{s1}, Figure S1: title, Table S1: title, Video S1: title.}


\paragraph{Acknowledgement} 
RZ is supported by an STFC Consolidated Grant, ST/P0000630/1.
I am grateful to  Saad Nabeebaccus and  Matt Rowe for careful reading of the notes and comments. Correspondence with  Matteo Di Carlo and Adrian Signer is further acknowledged. 
These notes were originally prepared for the EuroPLEx Summer School 2021 which fell victim to 
substantial shortening due to Covid. I intend to update these notes in the future with regards to structure dependence in the foreseeable future.

%% Optional
%\sampleavailability{Samples of the compounds ... are available from the authors.}

%%%%%%%%%%%%%%%%%%%%%%%%%%%%%%%%%%%%%%%%%%
%% Only for journal Encyclopedia
%\entrylink{The Link to this entry published on the encyclopedia platform.}

%%%%%%%%%%%%%%%%%%%%%%%%%%%%%%%%%%%%%%%%%%
%% Optional

%%%%%%%%%%%%%%%%%%%%%%%%%%%%%%%%%%%%%%%%%%
%% Optional
% Leave argument "no" if all appendix headings stay EMPTY (then no dot is printed after "Appendix A"). If the appendix sections contain a heading then change the argument to "yes".

\appendix

\section{Formal Aspects}

\subsection{The Low-theorem; a low-energy theorem}
\label{app:Low}

By the physical picture of  the multipole expansion of electrodynamics
a soft photon should be sensitive to the charge (monopole) and dipole distribution in the next    approximation. One  thus expects low energy theorems. In field theory such low energy theorems are connected 
 Ward identities. The circle of ideas closes as Ward identities derive from gauge invariance which in turn allows for the massless photon. 
Somewhat ``amusingly"   this theorem was put forward by a scientist to the name of Low and extended by others to what 
 is known as the Low-Burnett-Kroll-Goldberger-Gell-Mann  theorem 
\cite{Low:1954kd,Gell-Mann:1954wra,Low:1958sn,Burnett:1967km}.
The statement is that adding a real photon to a transition $\al \to \be$, the two first 
terms in an $E_\ga$-expansion are universal 
\begin{equation}
\label{eq:Low}
\matel{\be \ga(k,\la) }{ S} {\al} = ( J^{(0)}_\la +J^{(1)}_\la ) \matel{\be}{ S} {\al} + \ORD( E_\ga) \;,
\end{equation}
where the monopole and dipole term of $\ORD(E_\ga^{-1})$ and $\ORD(E_\ga^0)$ respectively, are given by 
\begin{equation}
 J^{(0)}_\la = \sum_{j} \hat{Q}_j \frac{ \eps^*(k,\la) \cdot \hat{p}_j    }{ k \cdot \hat{p}_j - i0}
 \;, \quad 
 J^{(1)}_\la = -i \sum_{j} \hat{Q}_j \frac{ \eps_\mu^*(k,\la)  k_\nu J_j^{\mu \nu}    }{ k \cdot \hat{p}_j - i0} \;.
\end{equation}
Above $J_j^{\mu\nu}  = i \hat{p}_j^{[\mu} \partial^{ \nu]}_{\hat{p}_j}$ is the orbital angular momentum operator and square brackets denoting antisymmetrisation in indices $a^{[\al}b^{\be]} = a^{\al}b^{\be}- a^{\be}b^{\al}$. 
Hatted quantities have the same meaning as described below \eqref{eq:leptonic-real}.  

The derivation is rather straightforward. Parameterising  the amplitude 
\begin{equation}
\matel{\be \ga(k,\la) }{ S} {\al} = \eps^*_\mu(k) {\cal A}^\mu(p_i,k) \;,
\end{equation}
with the additional convenient notation ${\cal A}_\mu(p_i,k)  \equiv  {\cal A}_\mu(\hat{p}_1 \dots \hat{p}_n,k) $   
which resolves the issue of in- and out-going states.
Now, \eqref{eq:Low} is obtained by making an ansatz for the  the Ward identity 
and solving it to the appropriate order.   We may write the ansatz as follows
\begin{equation}
\label{eq:ansatz}
{\cal A}_\mu(p_i,k)  = \sum_{j=1}^n \hat{Q}_j \frac{(p_j)_\mu}{k \cdot p_j} A_n(\hat{p}_1,\hat{p}_j+k,\hat{p}_n) + R_\mu({p}_i,k) \;,
\end{equation}
where $R$ stands for the remainder. The QED Ward identity reads
\begin{equation}
0  = k \cdot {\cal A}({p}_i,k)  = \sum_{j=1}^n \hat{Q}_j  A_n(\hat{p}_1,\hat{p}_j+k,\hat{p}_n) + k \cdot R(p_i,k) \;,
\end{equation}
and the essential step is to Taylor expand, as appropriate for a low-energy theorem, in $k$
\begin{equation}
0 = \sum_{j=1}^n (\hat{Q}_j  A_n|_{k=0} + k \cdot \partial_{\hat{p}_j}  A_n|_{k=0} )+ k \cdot R(p_i,k) + \ORD(k^2) \;.
\end{equation}
Note that, one will not be able to make a statement about $\ORD(k^2)$ as this probes the structure dependent part of the of the process. Equating terms one gets
\begin{alignat}{2}
& \ORD(|k|^0) : \quad & &  \sum_{j=1}^n \hat{Q}_j = 0 \;, \nonumber \\[0.1cm]
& \ORD(|k|^1) : \quad & &   \sum_{j=1}^n \hat{Q}_j  k \cdot \partial_{\hat{p}_j}  A_n|_{k=0} = -  k \cdot R \;,
\end{alignat} 
charge conservation at $ \ORD(|k|^0)$. 
To make the  $\ORD(|k|^1)$ equation useful one needs to uncontract the $k$. This is allowed 
since  no information is lost. This can be seen as follows. Assume $k \cdot y =0$ then the only non-trivial solution appears for two external vectors with $y_\mu = u \cdot k \, v_\mu -  v \cdot k \, u_\mu$. Hence
\begin{equation}
 R_\mu  = -  \sum_{j=1}^n \hat{Q}_j   (\partial_{\hat{p}_j} )_\mu A_n|_{k=0} \;.  
\end{equation}
Now we may take this equation and insert it into \eqref{eq:ansatz} and Taylor expand in $k$ 
to finally obtain
\begin{equation}
{\cal A}(p_j,k)_\mu = \sum_{j=1}^n \frac{\hat{Q}_j}{k \cdot p_j} L_j^\mu A_n(p_i) + \ORD(|k|) \;, \quad 
L_j^\mu = p_j^\mu - i k_\nu J^{\mu\nu}_j \;,
\end{equation}
 Low's theorem \eqref{eq:Low} in the notation used here. Low's theorem is believed to hold to all orders 
 and even non-perturbatively.  There are analogous soft-theorems for non-abelian gauge theories 
 and gravity but they only hold at tree level. 
 In (perturbative) non-abelian gauge theories loop corrections come with non-local terms  (presumably 
 since quarks and gluons are not physical particles) 
 which invalidate  the type of derivation followed above. However, the structure still shows some level of simplicity but is dependent case by case (e.g. \cite{Bern:2014vva} and pointers to the literature therein).

\subsection{KLN-theorem}
\label{app:KLN}

To what extent  QED with massless matter is well-defined is a question that was asked 
in the mid-sixties  by Kinosthita \cite{Kinoshita:1962ur} 
and Lee \& Nauenberg \cite{Lee:1964is}. Their work is known as the KLN-theorem: \emph{$S$-matrix elements squared are finite if one sums over 
energy-degenerate initial \emph{and}  final states}. 
Schematically
\begin{equation}
\label{eq:KLN}
\textrm{KLN-theorem:} \quad \sum_{i,f \in [E-\DEL,E+\DEL] } |\matel{f}{S}{i}|^2 = \textrm{finite} \;,
\end{equation}
which relies on unitarity and its proof involves the use of
time-ordered or old-fashioned perturbation theory. We refer the reader to Weinberg's book for an 
alternative proof closer to the coherent state approach \cite{Weinberg:1995mt}.
A few remarks might be helpful: 
\begin{itemize}
\item  If one sums over either all initial or all final states then the $S$-matrix elements
squared are of course finite:
$\sum_{i \textrm{ or } f } |\matel{f}{S}{i}|^2 = \textrm{finite}$ by 
unitarity ($S S^\dagger =1$ and $\mathbf{1} = \sum_{x} | x \rangle \langle x | $ for $x=i,f$) of the $S$-matrix.   
It is by selecting exclusive (final) states that IR-sensitivity appears. 
\item For QED, it turns out,  that summation over final states is sufficient for IR-finiteness, that is \eqref{eq:KLN} 
may be simplified to
$\sum_{f \in [E-\DEL,E+\DEL] } |\matel{f}{S}{i}|^2 = \textrm{finite} $ when all charged particles are massive. 
This goes hand in hand with the beforehand, known,  Bloch-Nordsieck mechanism 
which only demands summation over final states. 
QED is special in that the force carriers are not charged, unlike in QCD, and the soft photon Hilbert space can be decoupled such that only one sum is necessary
cf. chapter 13.4.  in \cite{Weinberg:1995mt}. To paraphrase Sterman \cite{Sterman:1993hfp}:
``From the viewpoint of the KLN-theorem the Block-Nordsieck mechanism seems somewhat accidental."
Historically the first counterexample to the Bloch-Nordsieck mechanism was found in QCD for 
$\bar q q \to \mu^+ \mu^- \bar q q$  at the 2-loop level \cite{Doria:1980ak}.

%\item One might be inclined to ask what the infinites mean in the context 
%of massless QED. The point is that no detector apparatus can distinguish 
%an electron from an electron with a photon at (nearly) zero angle.  Again the divergences are associated with an idealisation. For remarks on QCD and collider physics cf. \APP\ref{app:terminology} under IR safety.
\item The requirement of the summation over degenerate  energy  states does invalidate 
some differential decay rates or cross section as IR-safe and collinear-safe observables
(cf. the discussion in \SEC\ref{sec:semileptonic}). 
For IR-divergences caused by soft photons the 
standard explanation that their effects are not measurable, beyond some energy resolution $\DEL$, 
is after all rather satisfactory.  Collinear divergences in QCD, $\ln m_q$ (for $m_q \to 0$), indicate a problem in the formalism (cf. discussion in cf. \APP\ref{app:terminology} ) as quark masses are unphysical due 
to confinement.\footnote{The problem with QCD or non-abelian gauge theories, confined or not, is that  coloured states are not valid asymptotic states 
 since the colour of any state can always be changed by emitting 
a soft gluon, e.g. \cite{Smilga:2001ck}.}  
Collinear divergences in QED, $\ln m_e$ (for $m_e \to 0$), pose a whole different level 
of problems which presumably go into measurement problems and are beyond the authors's expertise.

\item The KLN-theorem is reminiscent of a theorem in CP-violation that states that if  one sums  over all final states that can rescatter into each other under the strong force, then the rate of particle and anti-particle process are the same 
($\sum_{f \in \textrm{rescatter}} (\Gamma(i \to f) - \Gamma(\bar i \to \bar f) = 0$) \cite{Bigi:2000yz}. 
Not only the flair of the theorem but also its method of proof, namely unitarity, is the same.
%\cite{Frye:2018xjj}
\end{itemize}

\subsection{Brief synopsis  of coherent states}
\label{app:coherent}

In this brief appendix we sketch  some elements  of coherent states following  the excellent exposition in 
the book on the conceptual framework of QFT 
\cite{Duncan}. Coherent states originally came from optics we refer the reader to the book 
on Quantum Optics \cite{gerry_knight_2004} for a thorough introduction.

One way to introduce the coherent 
state is as the state maximising the number-phase uncertainty relation. One can derive an analogue of the Heisenberg uncertainty relation $\Delta x \Delta p \geq \hbar$ for the particle number and the phase. Searching for a solution with $\Delta N = \Delta \phi$  and  justifiably truncating the Hilbert space, one arrives at the condition that this state is to 
be an eigenstate of the annihilation operator.\footnote{States where  
$\Delta N \neq \Delta \phi$ are known as squeezed states,  of interest in optics and described by minimal modifications only.}  This makes it clear that this state must be 
a coherent sum over the infinite series of all excitation modes.

Starting with the standard harmonic oscillator
($[a, a^\dagger]_{nm} = \de_{nm}$,  $a^\dagger \state{n} = \sqrt{n+1}\state{n+1}$, $a \state{0} = 0$)   and imposing the eigenvalue equation 
$a \state{\omega} = \omega \state{\omega}$  for a generic ansatz,  a set of recursion relation emerges which are solved to give the coherent state
\begin{equation}
\label{eq:coh}
a \state{\omega} = \omega \state{\omega} \;, \quad \state{\omega} = e^{-\frac{1}{2} |\omega|^2} \sum_{n \geq 0} \frac{\omega^n}{\sqrt{n!}} \state{n}  \;.
\end{equation}
Or alternatively 
\begin{equation}
\state{\omega}  = S(\omega) \state{0}  \;, \quad 
S(\omega) = e^{-\frac{1}{2} |\omega|^2}  e^{\omega a^\dagger} \;,
\end{equation}
 and 
the prefactor assures $\langle \omega | \omega \rangle =1$ 
(since $S(\omega)^\dagger  S(\omega) =1$).
This state saturates the  number-phase inequality and can  be regarded as a state close to a classical state.  For illustration let us consider an electromagnetic field in the $z$-direction  
with a single wave vector  $k$ (monochromatic). For its  corresponding  coherent state 
$| \omega  \rangle$ with unspecified polarisation, one has schematically 
\begin{equation}
\matel{ \omega }{E_z(x)}{\omega  } =  -2 e^{-\frac{1}{2} |\omega|^2} A \sin(k \cdot x + \theta) \;, \quad    \omega = A^{i \theta} \;,
\end{equation}
and this reveals the meaning of the eigenvalue $\omega$. Its radial part is the amplitude and its phase the phase shift. 
Of course in QFT there are infinitely many frequencies to which we turn further below. 
Moreover, note that  $|\vev{n|\omega}|^2 = e^{- |\omega|^2}\frac{\omega^n}{n!}$ follows a Poisson distribution. 
In the context of  QED each $n$ corresponds to the emission of $n$ undetectable soft photons. 
%In QED the prefactor $e^{-\frac{1}{2} |\omega|^2}$ is the analogue 
%of the (divergent) virtual amplitude.

For a (scalar) QFT , the analogue of the operator $S$ is given by
\begin{equation}
S(f) \propto \exp( \frac{1}{2} \int d^3 k (2\pi)^{3/2} \sqrt{2 E_k}  \tilde{f}(k) a^\dagger_k) \;,
\end{equation}
where $\tilde f$, Fourier transform of $f(x)$,  
is the momentum distribution defining the wave packet. 
 The state $f$ is then given by $\state{f} = S(f) \state{0}$
and the omitted normalisation factor ($e^{ - \frac{1}{2}|\omega|^2}$ in \eqref{eq:coh})  is the analogue of the virtual amplitude 
defined without  emission of extra soft-particles created by $a^\dagger$.  
The distribution $f(x)$ has a very direct in meaning in that it describes (normal ordered) expectation 
values, e.g.  for a real-valued field $\phi$:  $\matel{f}{\phi^n(x)}{f}/ \vev{f|f} = f(x)^n$, $\matel{f}{ (\partial_i \phi)^2(x)}{f}/ \vev{f|f} =(\partial_i f)^2(x)$ but  $\matel{f}{ \partial_0 \phi}{f} =0$). 

At last and crucially for this text, the coherent states used in the IR-definition of the S-matrix, corresponding to \eqref{eq:coherent}, is then implemented by  
$\tilde{f}(k) =  0 $ for $|\vec{k}| > \DEL  $.  This captures all photons 
with energies below $\DEL$ in the spirit of  \eqref{eq:coherent}.

\section{Some Practical Aspects of   QED in the Infrared}

\subsection{Infrared divergences at one-loop}
\label{app:regions}

We consider it worthwhile to briefly give the essence of how IR-singularities 
are identified in one-loop diagrams paralleling the real-emission discussion 
in \SEC\ref{sec:IR}.

The collinear divergences are simpler than the soft ones in the sense that one does 
not need to involve power counting arguments based on the dimension of spacetime. 
 collinear divergences occur when a massless particles is emitted from another massless particle
and the two momenta are collinear.  If either of the particles has a (small) mass $m$ then the divergence is regulated by $\ln m$ cf. \eqref{eq:basic}.

The soft-divergences are more subtle as the inverse power in the photon energy $E_\ga$. 
The criteria is that  two external momenta are to be on-shell with a photon propagating in the loop. The relevant power counting then assumes
\begin{equation}
  \int \frac{d^4 k }{k^2 (  (k+p_1)^2 - m_1^2)  (k+p_2)^2 - m_2^2)  }(1+ \ORD(k))  \to 
 \int \frac{dk k^3 d\Omega }{k^2 ( 2 k \cdot p_1 ) ( 2 k \cdot p_2 )     }  \;,
\end{equation}
where $p_{1,2}^2 = m_{1,2}^2$ and $\ORD(k)$ were dropped in the second step. We see that this integral is logarithmically divergent for $d=4$ when $k \to 0$
as previously stated.

There are algorithms to extract soft and collinear divergences at one-loop \cite{Dittmaier:2003bc} and 
two-loop \cite{Anastasiou:2018rib}.
An approach that works more generically is to realise that IR-singularities  are associated 
with singularities in the complex plane which in turn can be studied in perturbation theory 
by the Landau equations.  This involves though two further non-trivial steps. First 
one needs to check whether the singularity in question is on the first sheet. Second not every singularity or branch points leads to a IR-singularity. For example $(p^2-m^2) \ln (p^2-m^2)$ has a branch cut starting at $p^2 = m^2$  but is not singular at that point itself.
The second topic is discussed in detail in Sterman's book chapter 13 \cite{Sterman:1993hfp} as well as in 
his lecture notes \cite{Sterman:1995fz,Sterman:2004pd}.
The systematic development of singularities in terms of effective Lagrangians is ,the previously mentioned,
 soft-collinear effective theory \cite{Becher:2014oda} with the advantage of the systematic use of the equation of motion and a renormalisation group program.

 \subsection{How to handle non-analytic decay rates numerically}  
\label{app:pragmatic}

It seems relevant to briefly mention the practical problem of dealing with 
IR-divergences numerically. For the leptonic decay (in the point-like approximation) 
everything can be done analytically and then matters are straightforward. 
For the semileptonic case it is already more challenging but since there's just one
non-trivial phase space integral, namely when the photon couples to 
the pion and the lepton, it is still doable \cite{Ginsberg:1969jh}. 
In the generic case, if we take all particles to be charged  \cite{Isidori:2020acz}, 
it is maybe possible but the effort does not seem worthwhile. At higher loops in QCD this becomes totally 
unfeasible and people resort to so-called subtraction schemes (e.g. dipole, antenna or Catani-Seymour subtraction). The idea is simple, one decomposes  
\begin{equation}
\Ampone =  \Ampone_A +  (\Ampone- \Ampone_A ) \;,
\end{equation}
where $\Ampone_A$ is doable analytically and the term in bracket is free from IR-divergences. 
Preferably, it is also  free from large logs in order to avoid numerical 
instabilities.  It is for this reason that the evaluation of the phase space integral in  \eqref{eq:F02} in the leading log approximation is valuable in practice. 
It is fortunate that in this approximation both integrals can be shown to be separately Lorentz invariant!

\subsection{Terminology}
\label{app:terminology}
Whereas terminology can always be a hurdle for people learning a subject, QED corrections are riddled with multiple expressions meaning the same thing and 
are historic or context based rather than logical. This short appendix ought to help clarifying a few of these matters.
\begin{itemize}
\item When hadrons are treated as point-like particles one often refers to this approach as 
\emphB{scalar-QED} presumably in the context of scalar mesons such as the pion. Of course one can also treat a baryon as  point-like but it being a fermion then makes the term scalar-QED seem inappropriate. Going beyond the point-like approximation, resolving the 
hadrons beyond the monopole approximation, is referred to as a \emphB{structure-dependent} 
contribution which is the context of \SEC\ref{sec:beyond}.
\item IR-divergences are often synonymous with soft-divergences which includes soft-collinear divergences. Collinear terms, referred to as $\ln m_f$ in the text where $f$ stands for final states, are referred to as 
\emphB{collinear divergences}  if $m_f \to 0$ (when computing with massless quarks in QCD) or 
\emphB{(hard-)collinear logs} (if $m_f \ll m_i$).  Some authors refer to them as \emphB{mass-singularities} as well \cite{Muta:1998vi}.
It should usually be clear from the context but it is useful to be aware of the  potential  confusion.
\item The concept of \emphB{IR-safety} has been introduced by Sterman and Weinberg \cite{Sterman:1977wj} 
and means the following. An observable computed with quark and gluons  is IR-safe if
the quark masses can be taken to zero without encountering singularities (i.e. avoiding hard-collinear singularities of the $\ln m_q$-type). As previously stated, In the context of QCD this amounts to either defining inclusive enough quantities or 
legitimately absorbing collinear logs into hadronic objects  (jets or parton distribution functions) at the
expense of introducing a factorisation scale.  
\item In the context of $\ORD(\al)$ computations and the use of the Bloch-Nordsieck and KLN cancellations
of IR-divergences \eqref{eq:total-cancel} one refers to $\Gamma(i \to f)$ and $\Gamma(i \to f\ga) $ 
as the \emphB{non-radiative} and \emphB{radiative rate} respectively. Often the terms virtual and real are used synonymously since
those correspond to the precise $\ORD(\al)$-terms. 
\end{itemize}

%\externalbibliography{yes}
%\bibliographystyle{utphys}
\bibliography{../../Refs-dropbox/References_QED.bib}
\bibliographystyle{utphys}
%\bibliography{References_QED.bib}

\end{document}